\documentclass[11pt]{article}
\pdfoutput=1
\usepackage{amssymb}
\usepackage{amsmath}
\usepackage{amstext}
\usepackage{graphicx,epsfig}
\usepackage{epsfig}
\usepackage{verbatim} 
\usepackage{fancybox}
\usepackage{color}
\usepackage{ulem}
\usepackage{enumitem}
\usepackage{subfigure}
\usepackage{bbm}
\usepackage{parskip}
\usepackage{dsfont}
\usepackage[numbers,sort&compress]{natbib}
\usepackage{cancel}

\newcommand{\Comment}[1]{{}}
\definecolor{darkblue}{rgb}{0.15,0.35,0.55}
\definecolor{reddish}{rgb}{0.65, 0.2, 0.2}
\usepackage[linktocpage=true]{hyperref}
\hypersetup{
colorlinks=true,
citecolor=darkblue,
linkcolor=reddish,
urlcolor=darkblue,
pdfauthor={},
pdftitle={},
pdfsubject={}
}

\newcommand{\be}{\begin{equation}}
\newcommand{\ee}{\end{equation}}
\newcommand{\bea}{\begin{eqnarray}}
\newcommand{\eea}{\end{eqnarray}}
\newcommand{\beas}{\begin{eqnarray*}}
\newcommand{\eeas}{\end{eqnarray*}}
\newcommand{\nn}{\nonumber}

\def\({\left(}
\def\){\right)}

\def\gsim{ \lower .75ex \hbox{$\sim$} \llap{\raise .27ex \hbox{$>$}} }
\def\lsim{ \lower .75ex \hbox{$\sim$} \llap{\raise .27ex \hbox{$<$}} }

\DeclareMathOperator{\SO}{SO}

\def\sech{\mathop{\rm sech}\nolimits}
\def\csch{\mathop{\rm csch}\nolimits}

\title{}
\author{}

\numberwithin{equation}{section}

\begin{document}
%
~
\vspace{2truecm}
\begin{center}
{\LARGE \bf{Holographic CFTs on maximally symmetric spaces: correlators, integral transforms and applications}}
\end{center}

\vspace{1.3truecm}
\thispagestyle{empty}
\centerline{\Large Kurt Hinterbichler,${}^{\rm a}$ James Stokes${}^{\rm b}$ and Mark Trodden${}^{\rm b}$}
\vspace{.7cm}

\centerline{\it ${}^{\rm a}$Perimeter Institute for Theoretical Physics,}
\centerline{\it  31 Caroline St. N, Waterloo, Ontario, Canada, N2L 2Y5}

\vspace{.3cm}

\centerline{\it$^{\rm b}$Center for Particle Cosmology, Department of Physics and Astronomy,}
\centerline{\it University of Pennsylvania, Philadelphia, PA 19104, USA}

\vspace{.4cm}

\begin{abstract}
We study one and two point functions of conformal field theories on spaces of maximal symmetry with and without boundaries and investigate their spectral representations. Integral transforms are found, relating the spectral decomposition to renormalized position space correlators. Several applications are presented, including the holographic boundary CFTs as well as spacelike boundary CFTs,  which provide realizations of the pseudo-conformal universe.
\vspace{.03cm}
\noindent
\end{abstract}

\tableofcontents

\newpage

\section{Introduction}
Correlation functions in quantum field theories can present both short and long distance singularities. Short-distance singularities are regularization dependent.  They parametrize un-calculable high energy effects which are renormalized into the undetermined local couplings of the effective action\footnote{There are some exceptions, e.g. \cite{Closset:2012vp}.}.  Theories without a mass gap exhibit long-range correlations, which can lead to infra-red singularities in Fourier space.  These are calculable universal features, not dependent on regularization ambiguities or absorbable into local couplings.

In this paper we use a combination of quantum-field theoretic and holographic techniques to study the relationship between position and momentum space correlation functions in conformal field theories (CFTs) on maximally symmetric curved spaces, with and without boundaries.  We focus primarily on one and two-point functions, paying special attention to the short-distance  singularities and how they are to be renormalized into local counterterms.

Our general analysis encompasses the anti-de Sitter/boundary conformal field theory  (AdS/BCFT) correspondence \cite{Takayanagi:2011zk} (see \cite{Fujita:2011fp,Nozaki:2012qd} for reviews).  The AdS/BCFT correspondence can be regarded as a generalization of AdS/CFT \cite{Maldacena:1997re} to situations in which the dual field theory itself has some boundary or defect \cite{Cardy:2004hm}.  In this case, the bulk theory possesses a boundary $Q$ in addition to the usual asymptotic boundary $M$ of AdS$_{d+1}$. The intersection $\partial M = Q \cap M$ of the new boundary $Q$ with the CFT living on $M$ represents the defect or boundary of the CFT.  
In this case the dual field theory is called a boundary conformal field theory.
If the bulk boundary $Q$ is chosen to preserve some subgroup of the $\mathrm{O}(2,d)$ isometries of the bulk AdS$_{d+1}$, then the dual field theory is invariant under the corresponding subgroup of the conformal group.

There exist a number of existing examples of this general setup.  The metric for the Poincar\'{e} patch of AdS$_{d+1}$ is
\begin{equation}
	ds^2 = \frac{dz^2 - dt^2 + dx_1^2+\cdots+dx_{d-2}^2+dy^2}{z^2} \ ,\label{adsmetricp}
\end{equation}
where $z \in (0,\infty)$, and $(t,x_1,\ldots,x_{d-2},y)$ label the coordinates of the $d$-dimensional dual field theory at $z=0$.
The Randall-Sundrum or hard-wall AdS/QCD models \cite{Randall:1999ee,Erlich:2005qh} can be considered as an example, where the role of $Q$ is played by the IR brane which lies at a fixed value $z=z_\ast > 0$ of the Poincar\'{e} radial coordinate, and the role of $M$ is played by the $z=0$ boundary; $M=\mathbb{R}^{1,d-1}$ at $z\to 0$.  In this example $Q$ does not intersect $M$.  
Poincar\'{e} symmetry $\mathrm{ISO}(1,d-1)\subset \mathrm{O}(2,d)$ is respected but the dilation and special conformal symmetry of $\mathrm{O}(2,d)$ is  broken and this introduces a mass scale $1/z_\ast$ in the dual quantum field theory. The soft wall can be thought of as a generalization of the Randall-Sundrum I model \cite{Randall:1999ee}, which contains a back-reacting scalar field in the bulk. The scalar field becomes singular in the interior of AdS and forms a naked singularity which plays the role of the IR boundary brane.

Locally localized gravity \cite{Karch:2000ct} is another example where $Q$ is an AdS$_d$ submanifold of AdS$_{d+1}$ which intersects $M =\mathbb{R}^{1,d-1}$ along a flat, timelike surface $y = 0$.  This holographically realizes a CFT on a half space $y\in [0,\infty)$ whose boundary at $y = 0$ breaks the conformal group $\mathrm{O}(2,d)$ but leaves unbroken an $\mathrm{O}(2,d-1)$ subgroup.

If we instead take a suitable de Sitter submanifold $Q=\mathrm{dS}_{d}$, then we find that $Q$ intersects $M = \mathbb{R}^{1,d-1}$ on the flat spacelike surface $t = 0$ and the CFT is defined at times $t \in (-\infty,0]$. This is our proposal for the holographic dual to a new kind of conformal field theory which possesses a spacelike boundary at future infinity. These new CFTs find application in the pseudo-conformal universe scenario for early universe cosmology.   The pseudo-conformal universe \cite{Rubakov:2009np,Creminelli:2010ba,Hinterbichler:2011qk,Hinterbichler:2012mv} is an early universe scenario which serves as an alternative to inflation, in which the early universe is dominated by a CFT that spontaneously breaks the conformal group to a subgroup which is isomorphic to the group of de Sitter symmetries.   Here, contrary to most applications of AdS/CFT or dS/CFT to cosmology, the theory of cosmological interest is the boundary CFT.  Within this boundary CFT, there is a spacelike surface at $t=0$ which marks the point at which the pseudo-conformal phase ends and the universe must reheat and transition into a radiation dominated phase.  This spacelike surface is the boundary of the CFT, which makes it a wick-rotated version of a BCFT.   
 The boundary $t = 0$ now preserves a de Sitter subgroup $\mathrm{O}(1,d)\subset \mathrm{O}(2,d)$ and the most general vacuum expectation values for scalar operators of dimension $\Delta$ can evolve in time as $1/(-t)^{\Delta}$.  Our proposal can be considered as the hard-wall version of \cite{Hinterbichler:2014tka,Libanov:2014nla,Libanov:2015mha}.  

Our goal will be to study the simplest correlation function in these various situations.  Our initial motivation was to follow up on the analysis of \cite{Hinterbichler:2014tka} and holographically compute two-point functions in the conformal universe (which would correspond to the power spectra of interest in cosmology), but the results apply more widely to other BCFTs.

The organization is as follows. In Section \ref{sec:distn} we study two-point functions and their singularities in CFTs on flat space, the sphere, and hyperbolic space.  In Section \ref{sec:bcft} we review the construction of holographic BCFTs, including their one- and two-point functions from the gravity dual. We additionally present a new derivation of the AdS/BCFT two-point function which exploits the AdS slicing of the bulk and provides an additional test of the of the construction laid out in section \ref{sec:distn}. In section \ref{sec:pseudo} we provide additional calculations for the one-point and two-point function in the spacelike boundary (or pseudo-conformal) CFT. 

\section{CFT correlators on maximally symmetric spaces and their UV singularities}\label{sec:distn}

We would like to understand how two-point correlators, the treatment of their UV singularities, their interpretation as distributions, and their Fourier transforms, generalize to curved spaces. In particular, we consider maximally symmetric spaces, which have the same number of symmetries as flat $\mathbb{E}^d$ and are related to it by Weyl transformations. Physically, these spaces are solutions to the Einstein equations with a cosmological constant.  In this section, we will consider the cases of Euclidean CFTs on spaces without boundaries, moving on to cases with boundaries in Section \ref{sec:bcft}.

\subsection{Flat space\label{flatspacesubsn}}

We warm up by analyzing the singularity structure of the simplest possibility: a CFT on flat space without boundaries. We recall how local counter-terms must be introduced to remove the short-distance singularities of bare correlation functions. Our analysis differs from \cite{Osborn:1993cr} in that we employ cut-off, rather than differential regularization, which we found easier to generalize to curved spaces. We will see in particular examples how the renormalized correlation functions thus defined are implicitly determined in terms of their Fourier transform. 

Consider a CFT on flat $\mathbb{E}^d$ with $d \geq 3$. As is well known, the conformal symmetry fixes the form of the two-point correlator for scalar primary operators of dimension $\Delta$ to be $1/x^{2\Delta}$.  Naively, the Fourier transform of the function $1/x^{2\Delta}$ is generally ill-defined both in the UV and IR. The naive definition of the Fourier transform of the two-point function is
\begin{align}
	\tilde{G}_{\Delta,d}(k)
	& \stackrel{!}{=} \int d^d \vec{x} \, e^{- i \vec{k}\cdot\vec{x}} |\vec{x}|^{-2\Delta} \  \\
	& = V_{\mathbb{S}^{d-2}}\int_{0}^\pi d\theta (\sin\theta)^{d-2} \int_0^\infty dr \, r^{(d-2\Delta)-1} e^{- i r k \cos\theta} \ ,
\end{align}
with $V_{\mathbb{S}^{d-2}}$ the volume of the unit ${d-2}$ sphere.
We see that this integral is only convergent if
\begin{equation}
	\frac{d-2}{2} <\Delta < \frac{d}{2} \ ,\label{fakebound}
\end{equation} 
where the upper and lower bounds are UV and IR constraints, respectively. There exist plenty of CFTs with operators violating this naive bound\footnote{$\Delta\geq \frac{d-2}{2} $ is the unitarity bound for a scalar operator, saturated only for a free scalar, so the lower bound in \eqref{fakebound} would be violated only for a free scalar \cite{Mack:1975je}.}.

The above considerations underscore the well-known fact that correlation functions should be interpreted as distributions (generalized functions).   A distribution is a linear functional defined to act on some space of smooth test functions with nice prescribed behavior at infinity\footnote{Usually the space of test functions is taken to be the Schwartz space, the space of smooth functions which fall at infinity, along with any of its multiple derivatives, faster than any inverse power of the coordinates.  Unlike the space of functions with compact support, this has the advantage that the Fourier transform is always defined within the space of test functions. The corresponding distributions are known as tempered distributions.}.  The action of the correlation functional on a test function $f(x)$ is as follows
\begin{equation}\label{e:distribution}
\frac{1}{x^{2\Delta}}[f] = \int d^d x \frac{1}{x^{2\Delta}}f(x) \ .
\end{equation}
Due to the nice fall-off behavior of the test function,  this interpretation of the two-point correlator is free from IR divergences.  However it is still not defined because of possible UV divergences localized at $x=0$.  These are dealt with in the following way. We first define a regulated functional $\left.1/x^{2\Delta}\right|_\epsilon$ which is UV finite for $\epsilon > 0$.  There are many ways to do this.  One way, which we illustrate below, is to cut off the integral within some ball around the origin of radius $\epsilon$.  Another is differential regularization \cite{Freedman:1991tk} (reviewed in Appendix \ref{diffregappend}).  

Because the divergence is associated with the singularity at $x=0$, the divergent terms in \eqref{e:distribution} depend only on the value of the test function and its derivatives at $x=0$.  Because the divergences are localized, they can be cancelled by adding distributions which are delta functions and derivatives of delta functions, at the origin.
We define the renormalized two-point correlator as a distribution of the following form
\begin{equation}
	\langle \mathcal{O}_\Delta(x)\mathcal{O}_\Delta(0) \rangle_{\mathbb{E}^d}^{\rm ren} = \lim_{\epsilon \to 0}\left[\left.\frac{1}{x^{2\Delta}}\right|_{\epsilon} + c_1 \delta^d(x) + c_2 \square\delta^d(x) + \cdots\right] \ ,\label{finiterenorm}
\end{equation}
where the coefficients $c_1,c_2,\ldots$ are chosen to depend on $1/\epsilon$ in such a way that the result is finite as $\epsilon\rightarrow 0$ when~\eqref{finiterenorm} is integrated against an arbitrary test function.  

The infinite parts of the $c$'s are fixed by requiring finiteness, but the finite parts are undetermined and represent ambiguities that are not calculable from the theory.  Different regularization schemes will give different finite parts.  If we let $J$ be a source for the operator ${\cal O}$ and think in terms of the effective action $W[J]$ whose functional derivatives generate the correlators, these delta function ambiguities are precisely the local terms,
\be W[J]\supset \int d^dx\  c_1 J^2+c_2 J\square J+\cdots.\ee
The local terms are ambiguous and contribute only to correlators at equal points, whereas the non-local terms are finite and unambiguous and contribute to the correlators at separate points. 

For example, we can define the regulated functional by integrating only outside of a $d$-dimensional ball $B_\epsilon$ of radius $\epsilon$,
\begin{equation}
	\left.\frac{1}{x^{2\Delta}}\right|_{\epsilon}[f] = \int_{\mathbb{E}^d\backslash B_\epsilon} d^d x \frac{1}{x^{2\Delta}}f(x) \ ,
\end{equation}
in which case the coefficients $c_1,c_2,\ldots$ are either inverse powers or logarithms of $\epsilon$,
\begin{align}
	\left.\frac{1}{x^{2\Delta}}\right|_{\epsilon}[f]
		& = \int_{\epsilon}^\infty dr \, r^{d-1-2\Delta} \int d\Omega_{d-1} \left[f(0) + r\hat{x}^\mu \partial_\mu f(0) + \frac{r^2}{2}\hat{x}^\mu\hat{x}^\nu \partial_\mu \partial_\nu f(0) + \cdots \right] \ , \\
		& = V_{\mathbb{S}^{d-1}}\int_{\epsilon}^\infty dr \, r^{d-1-2\Delta} \left[f(0) + \frac{r^2}{2d} \square f(0) +\cdots  
		 \right] \ .
\end{align}
The set of divergences ends with a logarithm if $\Delta = d/2 + k$ ($k = 0,1,2,\ldots$). Taking $d = 4$ and $\Delta = 2$, for instance, we find
\begin{align}
	\left.\frac{1}{x^{4}}\right|_{\epsilon}[f]
		& = 2\pi^2 \int_{\epsilon}^\infty \frac{dr}{r} \left[f(0) + \mathcal{O}(r^2) \right] \ , \\
		& = -2\pi^2 \log (\mu\epsilon) f(0) + \mathrm{finite} \ .
\end{align}
The mass scale $\mu$ is arbitrary and ambiguous, because it can be changed by the addition of a finite local piece.  The renormalized two-point correlator is thus the following distribution
\begin{equation}
	\langle \mathcal{O}_2(x)\mathcal{O}_2(0) \rangle_{\mathbb{E}^4}^{\rm ren} = \lim_{\epsilon \to 0}\left[\left.\frac{1}{x^{4}}\right|_{\epsilon} +2\pi^2 \log(\mu\epsilon)\delta^4(x)\right] \ .
\end{equation}
This is finite and well-defined as a distribution, ambiguous only up to local delta contributions.  Note that in cases in which there is a logarithmic divergence, such as this one, the coefficient of the logarithm is unambiguous and calculable, and is responsible for violation of scale invariance at coincident points,
\begin{equation}
	\mu \frac{\partial}{\partial \mu}\langle \mathcal{O}_2(x)\mathcal{O}_2(0) \rangle_{\mathbb{E}^4}^{\rm ren} = 2\pi^2 \delta^4(x) \ .\label{scalevariationf}
\end{equation}
Cases without logarithmic divergences, for example a $\Delta=2$ operator in $d=3$,
\begin{align}\label{e:twopoint}
	\langle\mathcal{O}_2(x)\mathcal{O}_2(0)\rangle_{\mathbb{E}^3}^{\rm ren}
		& = \lim_{\epsilon \to 0}\left[\left.\frac{1}{x^4}\right|_{\epsilon} -\frac{4\pi}{\epsilon}\delta^3(x)\right] \ ,
\end{align}
do not exhibit scale-dependence at coincidence points.
Another important case which we will return to later is a marginal operator for which $\Delta=d$, for example $\Delta=3$ in $d=3$,
\begin{equation}
	\langle\mathcal{O}_3(x)\mathcal{O}_3(0)\rangle_{\mathbb{E}^3}^{\rm ren}
		= \lim_{\epsilon \to 0}\left[\left.\frac{1}{x^6}\right|_{\epsilon} - \frac{2\pi}{3}\left(\frac{2}{\epsilon^3}\delta^3(x) + \frac{1}{\epsilon}\square \delta^3(x)\right)\right] \ .
\end{equation}

Now consider the Fourier transform (the appropriate integral transform in flat space).  The ordinary Fourier transform of a test function $f$ is another test function $\tilde f$.  Given a distribution $G$, its Fourier transform is always defined and is the distribution $\tilde G$ which gives the same value acting on $\tilde f$ as $G$ does acting on $f$.    
By this definition, the Fourier transform $\tilde{G}_{\Delta,d}(k)$ of the renormalized two-point distribution $\langle \mathcal{O}_\Delta(x)\mathcal{O}_\Delta(0) \rangle_{\mathbb{E}^d}^{\rm ren}$ satisfies
\begin{equation}\label{e:fourierdist}
	\frac{1}{(2\pi)^d}\int d^d k \, \tilde{G}_{\Delta,d}(k)  \tilde{f}(k) = \int d^d x \, \langle \mathcal{O}_\Delta(x)\mathcal{O}_\Delta(0) \rangle_{\mathbb{E}^d}^{\rm ren}  f(x) \ .
\end{equation} 
It can be shown that $\tilde{G}_{\Delta,d}(k)$ is given by
\begin{equation}
	\tilde{G}_{\Delta,d}(k) = 2^{d-2\Delta}\pi^{d/2}\frac{\Gamma(d/2-\Delta)}{\Gamma(\Delta)} k^{2\Delta - d} \ , \label{e:analytic}
\end{equation}
when $\Delta \neq d/2 + k$, and contains terms logarithmic in $k$ otherwise \cite{Freedman:1991tk}.  We are free to add to this arbitrary polynomials in $k^2$, since these are the Fourier transforms of the ambiguous local contact terms.  Note that \eqref{e:analytic} is the expression which would be obtained by analytically continuing in $\Delta$ the Fourier transform from the region \eqref{fakebound} in which it is defined without distributional considerations.

As a concrete example, consider a Gaussian test function of some width $a>0$,
\begin{equation}
	\tilde{f}(k) = e^{- ak^2} \iff f(x) = \frac{e^{-x^2/(4a)}}{(2\sqrt{\pi a})^3} \ .
\end{equation}
In the example of $d=3$, $\Delta=2$ given above, the left-hand side of \eqref{e:fourierdist} trivially gives $-1/(4a^2)$ while the right-hand side evaluates to
\begin{equation}
	\mathrm{RHS} = \frac{1}{(2\sqrt{\pi a})^3}\lim_{\epsilon\to 0}\left[4\pi\int_{\epsilon}^\infty dr\frac{e^{-r^2/(4a)}}{r^2}-\frac{4\pi}{\epsilon}\right] = -\frac{1}{4a^2} \ .
\end{equation}

In summary, we must regulate the UV singularities in position-space correlators, e.g. by imposing some short-distance cut-off around coincident points. After renormalization, the resulting correlators are finite, with ambiguous finite contact terms, and are related to their spectral decompositions by the integral transform \eqref{e:fourierdist}.  IR divergences, on the other hand, are calculable and unambiguous (and can be physically important, e.g. \cite{Mermin:1966fe,Hohenberg:1967zz,Coleman:1973ci}) and are handled automatically by the distributional interpretation, requiring no special treatment.

\subsection{Sphere}

Next we consider a Euclidean CFT on the $d$-dimensional sphere $\mathbb{S}^d$, which is related by analytic continuation to a Lorentzian CFT on de Sitter space $\mathrm{dS}_d$. This example will prove to be important for understanding the pseudo-conformal universe.

The two-point function for a CFT on $\mathbb{S}^d$ can be found by exploiting the fact that the round sphere is related to flat space by a Weyl transformation. 
The Euclidean space metric in spherical coordinates, and the standard round metric on the sphere are
\begin{equation}
	ds_{\mathbb{E}^d}^2 = dr^2 + r^2 d\Omega_{d-1}^2, \quad\quad ds_{\mathbb{S}^d}^2 = d\theta^2 + \sin^2\theta d\Omega_{d-1}^2 \ .
\end{equation}
Consider the stereographic projection from $\mathbb{S}^d$ to $\mathbb{E}^d$, given by $r = \sin\theta/(1-\cos\theta)=\cot(\theta/2)$ and thus $dr = d\theta/(1-\cos\theta)$. Substituting, we find
\begin{equation}
	ds_{\mathbb{S}^d}^2 = (1-\cos\theta)^2 ds_{\mathbb{E}^d}^2 \ .
\end{equation}
Conformal field theory correlators transform under Weyl transformations (up to anomalies) as
\begin{equation}
	\langle \mathcal{O}_{\Delta_1}(x_1)\cdots\mathcal{O}_{\Delta_n}(x_n) \rangle_{\Omega^2 g} = \Omega(x_1)^{-\Delta_1} \cdots \Omega(x_n)^{-\Delta_n}\langle \mathcal{O}_{\Delta_1}(x_1)\cdots\mathcal{O}_{\Delta_n}(x_n) \rangle_g \ . \label{weylcorrela}
\end{equation}
Setting $\Omega = (1-\cos\theta)$ and using the known flat space form for the two-point function, we deduce the following bare two-point function on the sphere,
\begin{align}\label{e:sphere2pt}
	\langle \mathcal{O}_\Delta(\vec{n})\mathcal{O}_\Delta(\vec{n}') \rangle_{\mathbb{S}^d}
		& = \frac{1}{2^\Delta(1-\cos\Theta )^\Delta} \ ,
\end{align}
where
\begin{equation}
	\cos\Theta = \cos\theta\cos\theta' + \cos\alpha\sin\theta\sin\theta' \ .
\end{equation}
$\Theta$ is the geodesic distance between the two points in $\mathbb{S}^d$ and $\alpha$ is their angular separation in $\mathbb{S}^{d-1}$. It is noteworthy that the two-point function only depends on the geodesic distance between two points on the sphere, which follows from the symmetries of the problem. The normalization of $1/2^{\Delta}$ is such that the short-distance limit matches the normalization ${1/ x^{2\Delta}}$ for flat space.  

We now attempt to perform the analog of the Fourier transform, that is, expand the two-point distribution on the sphere into hyper-spherical harmonics
as
\begin{align}
	\langle\mathcal{O}_\Delta(\vec{n})\mathcal{O}_\Delta(\vec{n}') \rangle_{\mathbb{S}^d}^{\rm ren} 
		& = \sum_{l,\mathbf{m}} g_l \mathbb{Y}_{l\mathbf{m}}^\ast(\vec{n})\mathbb{Y}_{l\mathbf{m}}(\vec{n}') \ , \\
		& = \frac{1}{V_{\mathbb{S}^d}(d-1)}\sum_{l}(2l+d-1) g_l C_l^{(d-1)/2}(\vec{n}\cdot\vec{n}')  \ , \label{e:spheredecomp}
\end{align}
where we have used the addition theorem on the $d$-dimensional sphere,
\begin{equation}\label{e:sphereaddition}
	\sum_{\mathbf{m}} \mathbb{Y}_{l\mathbf{m}}^\ast(\vec{n})\mathbb{Y}_{l\mathbf{m}}(\vec{n}') = \frac{1}{V_{\mathbb{S}^d}(d-1)}\sum_{l}(2l+d-1)C_l^{(d-1)/2}(\vec{n}\cdot\vec{n}') \ ,
\end{equation}
and $C_l^\alpha(x)$ are the Gegenbauer polynomials defined by the generating function
\begin{equation}
\frac{1}{(1-2xt+t^2)^\alpha} = \sum_{l=0}^\infty C_l^\alpha(x)t^l \ .
\end{equation}
The coefficients $g_l$ are the analog of the Fourier transform.
The inverse of this transform allows us to calculate the $g_l$'s
\begin{align}
	g_l 
		& = \frac{1}{\mathbb{Y}_{l\mathbf{m}}(\vec{n}')}\int_{\mathbb{S}^d} d\vec{n} \, \mathbb{Y}_{l\mathbf{m}}(\vec{n}) \langle\mathcal{O}_\Delta(\vec{n})\mathcal{O}_\Delta(\vec{n}') \rangle_{\mathbb{S}^d}^{\rm ren} \ , \\
		& = \frac{V_{\mathbb{S}^{d-1}}}{C_l^{(d-1)/2}(1)}\int_{-1}^1 dx \, (1- x^2)^{(d-2)/2}\left[\frac{1}{2^\Delta(1-x)^\Delta} + \textrm{counter-terms} \right]C_l^{(d-1)/2}(x) \ , \label{gintegrandd}
\end{align}
where we have used rotational invariance to move $\vec{n}'$ to $\theta = 0$ and have also used that the expression is independent of $\mathbf{m}$ to set $\mathbf{m}=0$, in which case the spherical harmonics become proportional to Gegenbauer polynomials. 

As in the flat case, this integral transform is generally ill-defined unless counter-terms are included: the singularity of the integrand \eqref{gintegrandd} at $x=1$  leads to the non-physical bound
\begin{equation}
\Delta < \frac{d}{2} \ .
\end{equation}
This is easy to understand because the sphere is locally flat, so we expect the same UV divergences as \eqref{fakebound} on flat space. There is no lower bound, however, because the finite volume of the sphere naturally cuts off the IR divergence.

To study the UV singularity structure of the bare two-point correlator, we integrate it against a smooth test function on $\mathbb{S}^d \times \mathbb{S}^d$  of the form $f(\vec{n}\cdot \vec{n}')$ as follows,
\begin{align}
	\int_{\mathbb{S}^d} d\vec{n} \langle\mathcal{O}_\Delta(\vec{n})\mathcal{O}_\Delta(\vec{n}') \rangle_{\mathbb{S}^d} f(\vec{n}\cdot \vec{n}')
		& = V_{\mathbb{S}^{d-1}}2^{-\Delta}\int_{-1}^{1-\eta} dx (1-x^2)^{(d-2)/2}(1-x)^{-\Delta}f(x) \ , \notag \\
		& = V_{\mathbb{S}^{d-1}}2^{-\Delta}\int_{-1}^{1-\eta} dx (1-x^2)^{(d-2)/2}(1-x)^{-\Delta}\left[f(1) + (x-1) f'(1) +\cdots\right] \ ,
\end{align}
where $0<\eta \ll 1$ is a UV regulator, cutting off the region $x=1$ in the integral where the two points come together.   Expanding in powers of $1\over\eta$, there will be divergent parts which must be cancelled off by local counterterms.

For example, consider the case $\Delta =2$ and $d=3$, which has the divergent part.
\begin{align}
	\int_{\mathbb{S}^3} d\vec{n} \, \langle\mathcal{O}_2(\vec{n})\mathcal{O}_2(\vec{n}') \rangle_{\mathbb{S}^3} f(\vec{n}\cdot \vec{n}')
		& = \frac{2\pi\sqrt{2}}{\sqrt{\eta}}f(1) + \mathrm{finite} \ .
\end{align}
As in flat space, the divergence is local, depending only on the value of the test function at the point $x=1$ where the two points come together.  Subtracting off this divergence, the renormalized two-point correlator for an operator of this dimension should be defined as the distribution
\begin{equation}
	\langle \mathcal{O}_2(\vec{n})\mathcal{O}_2(\vec{n}') \rangle_{\mathbb{S}^3}^{\rm ren} = \lim_{\eta \to 0} \left[\frac{1}{2^2(1-\vec{n}\cdot\vec{n}')^2} - \frac{2\pi\sqrt{2}}{\sqrt{\eta}}\delta^3(\vec{n},\vec{n}')\right] \ ,
\end{equation}
where $\delta^d(\vec{n}_1,\vec{n}_2)$ is the covariant delta function on the sphere, defined such that
\begin{equation}
	\int_{\mathbb{S}^d} d\vec{n} \, \delta^3(\vec{n},\vec{n}')f(\vec{n}\cdot\vec{n}') = f(1) \ .
\end{equation}
In terms of the $x = \cos\Theta$ coordinate,
\begin{equation}
	\delta^d(\vec{n},\vec{n}') = \frac{\delta(1-x)}{V_{\mathbb{S}^{d-1}}(1-x)^{(d-2)/2}} \ .
\end{equation}
Expanding for small $\Theta$ we obtain
\begin{equation}
	\delta^d(\vec{n},\vec{n}') \sim \frac{\delta(\Theta)}{V_{\mathbb{S}^{d-1}}\Theta^{d-1}} \sim \delta^d(x) \ ,
\end{equation}
where $\delta^d(x)$ is the delta function in flat space. Let us check that the short distance behavior of this correlator agrees with flat space. We have $\eta = 1 - \cos\epsilon \sim  \epsilon^2/2$ and thus we reproduce \eqref{e:twopoint},
\begin{equation}
	\langle \mathcal{O}_2(\vec{n})\mathcal{O}_2(\vec{n}') \rangle_{\mathbb{S}^3}^{\rm ren} \sim \lim_{\epsilon \to 0} \left[\frac{1}{x^4} - \frac{4\pi}{\epsilon}\delta^3(x)\right] \ ,
\end{equation}
where $x$ is now the physical distance between $\vec{n}$ and $\vec{n}'$ and $\epsilon$ is the physical cut-off distance. Now let us calculate the $g_l$'s for our renormalized correlation function.  Since we have a well defined distribution, the integral transform should exist and hence the $g_l$'s will be finite.
We get
\begin{align}
	g_l
		& = \frac{1}{\mathbb{Y}_{l\mathbf{m}}(\vec{n}')}\int_{\mathbb{S}^3} d\vec{n} \, \mathbb{Y}_{l\mathbf{m}}(\vec{n}) \langle\mathcal{O}_2(\vec{n})\mathcal{O}_2(\vec{n}') \rangle_{\mathbb{S}^3}^{\rm ren} \ , \\
		& = \lim_{\eta \to 0}\frac{1}{\mathbb{Y}_{l\mathbf{m}}(\vec{n}')}\int_{\mathbb{S}^3} d\vec{n} \, \mathbb{Y}_{l\mathbf{m}}(\vec{n}) \left[\frac{1}{2^2(1-\vec{n}\cdot\vec{n}')^2} - \frac{2\pi\sqrt{2}}{\sqrt{\eta}}\delta^3(\vec{n},\vec{n}')\right] \ , \\
		& = \lim_{\eta \to 0}\left[- \frac{2\pi\sqrt{2}}{\sqrt{\eta}}+ \frac{V_{\mathbb{S}^{2}}}{U_l(1)}\int_{-1}^{1-\eta} dx (1- x^2)^{1/2}\frac{1}{2^{2}(1-x)^{2}}U_l(x)\right] \ . \label{e:glren}
\end{align}

If instead we define $g_l$ by analytic continuation in $\Delta$ from the region in which \eqref{gintegrandd} is defined, we obtain\footnote{See also \cite{Gubser:2002vv}, which is missing a factor of $2^{-\Delta}$.} 
\begin{align}
	g_l
		& = \frac{\pi^{d/2}}{2^{2\Delta - d}}\frac{\Gamma(d/2-\Delta)}{\Gamma(\Delta)}\frac{\Gamma(l + \Delta)}{\Gamma(d + l - \Delta)} \ ,  \label{e:gl}
\end{align}
where $\Delta \neq d/2 + k$ ($k=0,1,2,\ldots$). 
One can check by direct evaluation with $l=0,1,2,\ldots$ that \eqref{e:gl} agrees with the formula \eqref{e:glren} obtained by properly renormalizing, namely 
\begin{equation}
	g_l = -\pi^2(l+1) \ .
\end{equation}
We see that analytic continuation in $\Delta$ corresponds to minimal subtraction in the hard cut-off formalism, as was the case on flat space. The spectral decomposition \eqref{e:gl} is thus related to the renormalized two-point function by the following integral transform,
\begin{align}
	\sum_{l,\mathbf{m}}g_l f_l^\ast  \, \mathbb{Y}_{l\mathbf{m}}(\vec{n}')\mathbb{Y}_{l\mathbf{m}}^\ast(\vec{n}'')
		&  = \int_{\mathbb{S}^d} d\vec{n} \langle\mathcal{O}_\Delta(\vec{n})\mathcal{O}_\Delta(\vec{n}') \rangle_{\mathbb{S}^d}^{\rm ren} f(\vec{n}\cdot\vec{n}'') \ , 
\end{align}
or, written in terms of Gegenbauer polynomials,
\begin{align}\label{e:spheredistn}
	\frac{1}{V_{\mathbb{S}^{d}}(d-1)}\sum_l g_l f_l^\ast (2l+d-1)C_l^{(d-1)/2}(\vec{n}'\cdot \vec{n}'')
		&  =  \int_{\mathbb{S}^d} d\vec{n} \langle\mathcal{O}_\Delta(\vec{n})\mathcal{O}_\Delta(\vec{n}') \rangle_{\mathbb{S}^d}^{\rm ren} f(\vec{n}\cdot\vec{n}'') \ .
\end{align}
The above formula is the analog of the flat-space Fourier transform \eqref{e:fourierdist}.

To further illustrate, consider a gaussian test function on the sphere.  We can make a gaussian on the sphere by stereographically mapping a gaussian on $\mathbb{E}^d$ to the sphere. Starting with the smooth test function $e^{-r^2}$ on $\mathbb{E}^d$ (with $r$ the polar radial coordinate) we obtain the following smooth test function on $\mathbb{S}^d$,
\begin{equation}
	f(\vec{n}\cdot\vec{n}'') = \exp\left(-\frac{1+\vec{n}\cdot\vec{n}''}{1-\vec{n}\cdot\vec{n}''}\right) \ ,
\end{equation}
where we have made the following identifications,
\begin{equation}
	r = \cot(\theta/2), \quad \quad \vec{n} \cdot\vec{n}'' = \cos\theta \ .
\end{equation}
Computing the corresponding $f_l$'s gives
\begin{align}
	f_l 
		& = \frac{1}{\mathbb{Y}_{l\mathbf{m}}(\vec{n}')}\int_{\mathbb{S}^3} d\vec{n} \, \mathbb{Y}_{l\mathbf{m}}(\vec{n}) f(\vec{n}\cdot\vec{n}') \ , \\
		& = \frac{4\pi}{U_l(1)} \int_{-1}^1 dx (1-x^2)^{1/2}U_l(x)\exp\left(-\frac{1+x}{1-x}\right) \ .
\end{align}
It is convenient to choose $\vec{n}'' \cdot \vec{n}' = - 1$ so that the Gaussian is peaked when the arguments of the two-point distribution coincide. We then obtain for the right-hand side of \eqref{e:spheredistn},
\begin{align}
\textrm{RHS} 	
	& = 4\pi\lim_{\eta \to 0}\int_{-1}^1dx(1-x^2)^{1/2}\left[\frac{1}{2^2(1-\vec{n}\cdot\vec{n}')^2} - \frac{2\pi\sqrt{2}}{\sqrt{\eta}}\frac{\delta(1-x)}{4\pi(1-x^2)^{1/2}}\right]\exp\left(-\frac{1-x}{1+x}\right) \ , \notag \\
	& = 4\pi\lim_{\eta \to 0}\int_{-1}^{1-\eta}dx(1-x^2)^{1/2}\frac{\exp\left(-\frac{1-x}{1+x}\right)}{2^2(1-x)^2} - \frac{2\pi\sqrt{2}}{\sqrt{\eta}} \ , 
\end{align}
while on the left-hand side we obtain an infinite sum over Chebyshev polynomials
\begin{equation}
	\textrm{LHS} = -\frac{1}{2}\sum_{l=0}^\infty f_l^\ast (l+1)^2 U_l(-1) \ .
\end{equation}
Numerically computing the $f_l$'s it is easy to see that that LHS and RHS agree, as they should.

\subsection{Hyperboloid}

The next example we treat is the hyperboloid CFT (see e.g. \cite{Aharony:2010ay}), where we will see that correlators continue from the sphere in a simple way by analytic continuation of the angular momentum to complex values, as in \cite{Polyakov:2014rfa}.

Analytically continuing the sphere $\mathbb{S}^d$ to negative curvature we obtain the $d$-dimensional hyperbolic space $\mathbb{H}_d$, which is the Euclidean continuation of anti de Sitter space AdS$_d$. 
The analysis for the hyperboloid CFT proceeds similarly to the sphere. The conformal map from $\mathbb{E}^d$ to $\mathbb{H}_d$ is given by $r = \coth(\rho/2) = \sinh\rho/(\cosh\rho-1)$,
\begin{equation}
	ds_{\mathbb{H}_d}^2  = d\rho^2 + \sinh^2 \rho d\Omega_{d-1}^2 = (1-\cosh\rho)^2 (dr^2 + r^2 d\Omega_{d-1}^2) \ ,
\end{equation}
and hence the conformal factor is  $\Omega=(1-\cosh\rho)$. Using \eqref{weylcorrela} and the known flat space form for the two-point function, it follows that the bare two-point function on the hyperboloid is given by 
\begin{align}
	\langle\mathcal{O}_\Delta(n)\mathcal{O}_\Delta(n') \rangle_{\mathbb{H}_d}
		& = \frac{1}{2^\Delta(\cosh \ell-1)^\Delta} \ ,
\end{align}
where
\begin{equation}
	\cosh\ell = \cosh\rho\cosh\rho' - \cos\alpha\sinh\rho\sinh\rho' \ .
\end{equation}
$\ell$ is the geodesic distance on $\mathbb{H}_d$, and $\alpha$ is the angular separation of the two points in $\mathbb{S}^{d-1}$.  Expanding the two-point distribution into eigenfunctions $\psi_{p,l,\mathbf{m}}$ of the Laplacian on $\mathbb{H}_{d}$ (reviewed in appendix \ref{app:eigen}) gives
\begin{align}
	\langle\mathcal{O}_\Delta(n)\mathcal{O}_\Delta(n') \rangle_{\mathbb{H}_d}^{\rm ren}
		& = \int_0^\infty dp \, g(p) \sum_{l,\mathbf{m}} \psi_{p,l,\mathbf{m}}(n)\psi_{p,l,\mathbf{m}}(n')^\ast \ .
\end{align}
The right-hand side can be expressed in terms of the geodesic distance between $n$ and $n'$ with the help of the addition theorem \cite{bander}
\begin{equation}
	\sum_{l,\mathbf{m}}\psi_{p,l\mathbf{m}}(n) \psi_{p,l\mathbf{m}}^\ast(n') = \frac{1}{2\pi}(2\pi\sinh \ell)^{(2-d)/2} \left|\frac{\Gamma((d-1)/2+ip )}{\Gamma(ip)} \right|^2 P_{-1/2+ip}^{(2-d)/2}(\cosh \ell) \ , \label{e:addition}
\end{equation}
where $n\cdot n' = \cosh\ell$. We will focus on the case when $d$ is odd for simplicity, since in this case the Legendre functions can be expressed in terms of Gegenbauer functions. The generalization to even $d$ is straightforward. The addition theorem for $d$ odd is
\begin{equation}
	\sum_{l,\mathbf{m}}\psi_{p,l\mathbf{m}}(n) \psi_{p,l\mathbf{m}}^\ast(n') = \frac{2i}{V_{\mathbb{S}^d}(d-1)} p \, C^{(d-1)/2}_{-(d-1)/2 +ip}(\cosh \ell) \ ,
\end{equation}
and thus
\begin{align}
	\langle\mathcal{O}_\Delta(n)\mathcal{O}_\Delta(n') \rangle_{\mathbb{H}_d}^{\rm ren}
		& = \frac{2i}{V_{\mathbb{S}^{d}}(d-1)} \int_0^\infty dp \, p \, g(p) \, C^{(d-1)/2}_{-(d-1)/2 +ip}(\cosh \ell) \ .
\end{align}
Hence, we see that, at least in the case of odd dimension $d$, there is a simple relationship between the spectral decomposition of the two-point function on the sphere and the hyperboloid; namely, we simply take the corresponding expression on the sphere \eqref{e:spheredecomp} and analytically continue the angular momentum quantum number to complex values, corresponding to the principal series of unitary irreducible representations of $\SO(1,d)$ \cite{Vil}
\begin{equation}
	l = -\frac{d-1}{2}+ip \ , \quad \quad p \geq 0 \ .
\end{equation}
The spectral decomposition can be inverted to give
\begin{align}
	g(p) 
		& = \frac{1}{\psi_{p,l,\mathbf{m}}(n)} \int dn \, \psi_{p,l,\mathbf{m}}(n')\langle\mathcal{O}_\Delta(n)\mathcal{O}_\Delta(n') \rangle_{\mathbb{H}_d}^{\rm ren} \ , \\
		& = \frac{V_{\mathbb{S}^{d-1}}}{C_{-(d-1)/2+ip}^{(d-1)/2}(1)}\int_{1}^\infty dz (z^2- 1)^{(d-2)/2}\left[\frac{1}{2^{\Delta}(z-1)^{\Delta}}+ \textrm{counter-terms}\right]C_{-(d-1)/2+ip}^{(d-1)/2}(z) \ .
\end{align}
Here we have used that the expression is independent of $l$ and $\mathbf{m}$ to set them both to zero. This allows us to make use of the following identity which expresses the wavefunctions in terms of Gegenbauer functions
\begin{equation}
	C_{-(d-1)/2+ip}^{(d-1)/2}(\cosh \ell) = (\sinh \ell)^{(d-2)/2} P^{(2-d)/2}_{-1/2+ip}(\cosh \ell) \frac{2^{(d-2)/2}\Gamma(d/2)\Gamma((d-1)/2+ip)}{\Gamma(d-1)\Gamma(ip - d/2 + 3/2)} \ .
\end{equation}
That is,
\begin{equation}
	\psi_{p,0,0}(r,\Omega) 
		\propto \frac{\Gamma((d-1)/2 + ip)}{\Gamma(ip)} (\sinh r)^{(2-d)/2}P_{-1/2+ip}^{(2-d)/2}(\cosh r) 
		\propto C_{-(d-1)/2+ip}^{(d-1)/2}(\cosh r) \ .
\end{equation}

The generalization of the integral transformation \eqref{e:fourierdist} is now
\begin{equation}
	\int_0^\infty dp\sum_{l,\mathbf{m}}g(p) f(p)^\ast  \, \psi_{p,l,\mathbf{m}}(\vec{n}')\psi_{p,l,\mathbf{m}}(\vec{n}')^\ast = \int_{\mathbb{H}_d} dn \langle\mathcal{O}(n)\mathcal{O}(n') \rangle_{\mathbb{H}_d}^{\rm ren} f(n \cdot n') \ ,
\end{equation}
where
\begin{equation}
	f(p) = \frac{V_{\mathbb{S}^{d-1}}}{C_{-(d-1)/2+ip}^{(d-1)/2}(1)}\int_{1}^\infty dz (z^2- 1)^{(d-2)/2}f(z)C_{-(d-1)/2+ip}^{(d-1)/2}(z) \ . \label{e:fspectral}
\end{equation}
Using the addition theorem this becomes simply
\begin{equation}\label{e:hyperbolicdist}
	\frac{2i}{V_{\mathbb{S}^d}(d-1)} \int_{0}^\infty dp \, p \, C^{(d-1)/2}_{-(d-1)/2 +ip}(1) g(p) f(p)^\ast = \int_{\mathbb{H}_d} dn \langle\mathcal{O}(n)\mathcal{O}(n') \rangle_{\mathbb{H}_d}^{\rm ren} f(n \cdot n') \ .
\end{equation}

Let us test this formula by focusing on the case of a $\Delta = 2$ scalar operator in $d=3$ dimensions. Following the same steps as on the sphere we obtain the renormalized two-point correlator
\begin{align}
	\langle \mathcal{O}_2(n)\mathcal{O}_2(n') \rangle_{\mathbb{H}_3}^{\rm ren}
		& = \lim_{\eta \to 0} \left[\frac{1}{2^2(n\cdot n'-1)^2} - \frac{2\pi\sqrt{2}}{\sqrt{\eta}}\delta^3(n,n')\right] \ , \\
	g(p)
		& = \lim_{\eta \to 0}\left[- \frac{2\pi\sqrt{2}}{\sqrt{\eta}}+ \frac{V_{\mathbb{S}^{2}}}{U_{-1+ip}(1)}\int_{1+\eta}^{\infty} dz (z^2-1)^{1/2}\frac{1}{2^{2}(z-1)^{2}}U_{-1+ip}(z)\right] \ . \label{e:gspectral}
\end{align}
As before, consider a Gaussian test function to illustrate.  We recall that $\ell$ is related to $z$ by the relation $z=\cosh\ell$, so the natural analog of a Gaussian on the hyperboloid is the test function $f(z) = e^{-z}$. 
In order to evaluate the integral on the left-hand side of \eqref{e:hyperbolicdist} we need the spectral representations of $\langle\mathcal{O}_2(n)\mathcal{O}_2(n') \rangle_{\mathbb{H}_3}$ and $f(\vec{n}\cdot\vec{n}')$ which are given by \eqref{e:gspectral} and \eqref{e:fspectral}, respectively. The integral defining $f(p)$ was evaluated numerically for different values of $p$ and numerically interpolated. The integral defining $g(p)$, while difficult to evaluate, can be guessed by analytical continuation from the sphere.  Substituting $l= -(d-1)/2 + ip$, dropping a factor of $i$ and multiplying by a measure factor of $\coth(\pi p)$ one finds agreement with the numerics. Substituting the approximate expression for $f(p)$ and the exact expression for $g(p)$ into the left-hand side of \eqref{e:hyperbolicdist} and carrying out the final $p$-integral numerically leads to excellent agreement. Here we demonstrate the numerics for a $\Delta = 2$ operator in $d=3$ dimensions,
\begin{align}
	\int_{\mathbb{H}_3} dn \langle\mathcal{O}_2(n)\mathcal{O}_2(n') \rangle_{\mathbb{H}_3}^{\rm ren} f(n \cdot n')
		& = \lim_{\eta\to 0}\left[4\pi\int_{1+\eta}^\infty dz(z^2-1)^{1/2}\frac{e^{-z}}{2^2(z-1)^2}-\frac{2\pi\sqrt{2}}{\sqrt{\eta}}e^{-1}\right] \ , \\
		& \simeq -5.118 \ . \\
	\frac{i}{2\pi^2} \int_{0}^\infty dp \, p \, U_{-1 +ip}(1) g(p) f(p)^\ast 
		& \simeq 5.118 \ .
\end{align}
\begin{figure}[h]
\centering
\includegraphics[width=70mm]{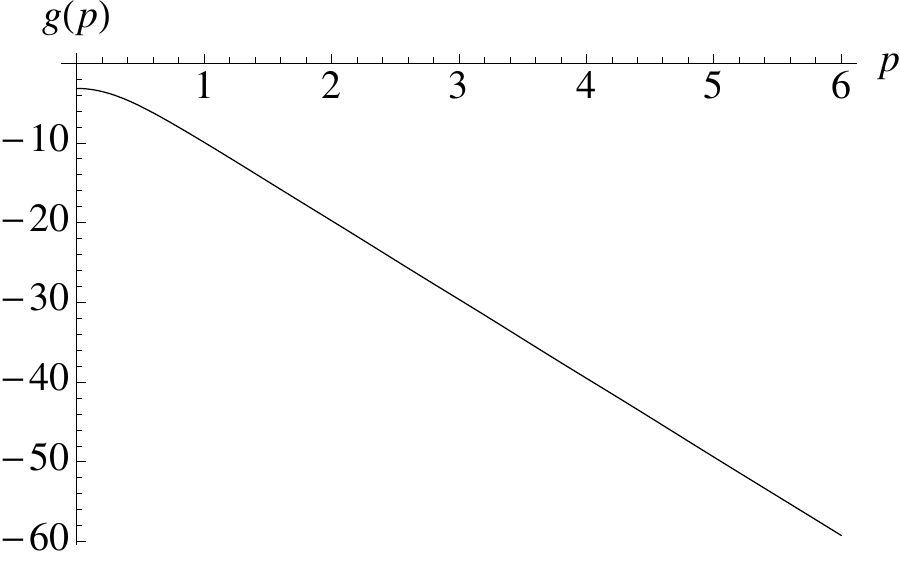}\includegraphics[width=70mm]{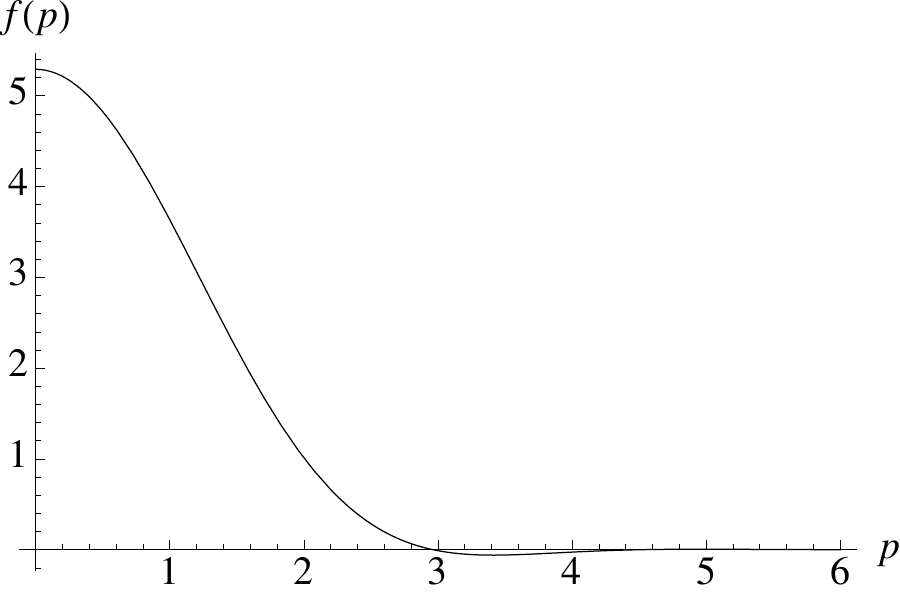}
\caption{Spectral representations $\langle\mathcal{O}_2(n)\mathcal{O}_2(n') \rangle_{\mathbb{H}_3}$ (left) and $f(\vec{n}\cdot\vec{n}')$ (right) obtained by numerical interpolation.}
\label{fig:interpolations}
\end{figure}
Let us now consider a marginal operator $\Delta = d = 3$. We have
\begin{equation}
	\frac{i}{2\pi^2} \int_{0}^\infty dp \, p \, U_{-1 +ip}(1) g(p) f(p)^\ast \simeq 1.1659 \ , \quad \quad g(p) = \frac{\pi^2}{12}p(1+p^2)\coth(\pi p) \ ,
\end{equation}
and
\begin{align}
	\int_{\mathbb{H}_3} dn \langle\mathcal{O}_3(n)\mathcal{O}_3(n') \rangle_{\mathbb{H}_3}^{\rm ren} f(n \cdot n')
		& = \lim_{\eta\to 0}\left[4\pi\int_{1+\eta}^\infty dz(z^2-1)^{1/2}\frac{e^{-z}}{2^3(z-1)^3}-\frac{\pi\sqrt{2}}{3\eta^{3/2}}e^{-1} + \frac{3\pi}{2\sqrt{2\eta}}e^{-1}\right] \ , \notag \\
		& \simeq 1.1659 \ .
\end{align}
\section{Holographic Boundary CFT}\label{sec:bcft}
In this section we will review the calculation of the one- and two-point functions for a boundary CFT from holography and then connect this with the integral transforms of the previous section. We begin with the AdS$_{d+1}$ metric \eqref{adsmetricp} and change to radial coordinates in the $z$ and $y$ variables: $(z,y) = (\eta\cos\phi,\eta\sin\phi)$.  We will also use the coordinate $\rho$, defined by $(z,y) = (\eta \sech \rho, \eta \tanh\rho)$.  The metric in these coordinates is
\begin{align}
	ds^2
		& = \frac{1}{\cos^2\phi}(d\phi^2 + ds_{AdS_d}^2) \  \\
		& = d\rho^2 + \cosh^2\rho \, ds^2_{AdS_d} \ .
\end{align}
This covers the full Poincar\'{e} patch of AdS$_{d+1}$ if $-\infty <\rho < \infty$, or $-\pi/2 < \phi <\pi/2$ (the UV boundary is at $\rho = -\infty$, or $\phi=-\pi/2$). The claim of the AdS/BCFT correspondence is that we obtain the holographic dual to a half-space CFT by restricting $ - \infty <\rho < \rho_\ast$ for some $\rho_\ast$. This effectively cuts the space off in the IR and introduces a second boundary $Q$ at $\rho_\ast$ in addition to the usual UV boundary at $\rho=-\infty$. The surface $Q$ defined by $\rho = \rho_\ast$ is given in Poincar\'{e} coordinates by $z=\eta\sech\rho_\ast$ and $y = \eta\tanh\rho_\ast$. Hence, $Q$ is defined by a curve in the $(y,z)$ plane
\begin{equation}
	y = z \sinh\rho_\ast \ .
\end{equation}
Notice that if we choose $\rho_\ast = 0$ then $Q$ is given simply by $y = 0$. For general $\rho_\ast$ let us define $\tan\theta = \sinh\rho_\ast$ and consider the rotation
\begin{equation}
	\begin{pmatrix}\tilde{y} \\ \tilde{z}\end{pmatrix} = \begin{pmatrix} \cos\theta & -\sin\theta \\ \sin\theta & \cos\theta \end{pmatrix}\begin{pmatrix}y \\ z\end{pmatrix} \ .
\end{equation}
Now we see that $Q$ lies at $\tilde{y} = 0$. 

Assume that localized on $Q$ there is a linear coupling 
\begin{equation}
	S_Q = \int_Q d^4 x\sqrt{h} \, a \phi \ ,\label{boundarytermss}
\end{equation}
with $\sqrt{h}$ the induced volume form on $Q$ and $a$ a constant. It is natural to add such a term because, from a Witten diagram point of view, it can be seen to correspond to giving a vacuum expectation value to the dual operator \cite{DeWolfe:2001pq}. An alternative possibility would be to add a boundary mass term $\int_Q d^d x \sqrt{h} \, b\,\phi^2$. Taking $b\to\infty$ realizes the Dirichlet boundary condition on $Q$.
The variation of this term does not affect the bulk equations of motion but contributes to the boundary variation.  The total boundary variation is
\begin{equation}
	\delta S = \int_Q d^4 x \sqrt{h} \delta\phi (n^\mu \partial_\mu \phi + a ) \ , \label{boundaryvarr}
\end{equation}
where $n^\mu$ is the unit normal to $Q$.  The first term is the boundary term coming from the variation of the bulk kinetic term after integration by parts, and the second term comes from varying \eqref{boundarytermss}.
The variational principle requires \eqref{boundaryvarr} to vanish for arbitrary $\delta\phi$, which requires the boundary condition on $Q$ to be of Neumann form
\begin{equation}
	(n^\mu \partial_\mu\phi  + a)|_Q = 0 \ .
\end{equation}

In our case, we have $n_\mu dx^\mu = c \, d\tilde{y}$ where $c$ is determined by the normalization condition $n_\mu n^\mu = 1$ or $g^{\tilde{y}\tilde{y}}c^2=1$. The metric written in terms of these variables is
\begin{equation}
	ds^2 = \frac{d\tilde{z}^2 - dt^2 + dx_1^2+dx_2^2+d\tilde{y}^2}{z^2} = \frac{d\tilde{z}^2 - dt^2 + dx_1^2+dx_2^2+d\tilde{y}^2}{(\tilde{z}\cos\theta - \tilde{y}\sin\theta)^2} \ .
\end{equation}
Hence, $g^{\tilde{y}\tilde{y}} = z^2$, $c = \pm 1/z$, $n^\mu \partial_\mu = c g^{\tilde{y}\tilde{y}}\partial_{\tilde{y}} = \pm z \partial_{\tilde y}$. Choosing the plus sign we obtain the boundary condition
\begin{align}
	\left.\partial_{\tilde{y}}\phi\right|_{\tilde{y} = 0} + \frac{a}{\tilde{z}\cos\theta}
		& = 0 \ , \\
	\left.(\cos\theta \partial_y - \sin\theta \partial_z)\phi\right|_{y = z\tan\theta} + \frac{a}{z}
		& = 0 \ .
\end{align}

\subsection{One-point function}
Let us consider the Fourier transform of the field configuration in the $y$-direction,
\begin{equation}
	\phi(y,z) = z^{d/2}\int_{-\infty}^{\infty} dq \, f_q(z) c(q) e^{i q y} \ .
\end{equation}
Substituting into the scalar equation in the flat slicing
\begin{equation}
	z^2 \partial_z^2 \phi - (d-1) z \partial_z \phi + z^2 \eta^{\mu\nu}\partial_\mu\partial_\nu\phi
		= m^2 \phi
\end{equation}
we obtain
\begin{equation}
	z^{d/2+2}\int_{-\infty}^{\infty} dq\,f_q(z)c(q)e^{iqy}\left[\frac{f_q''(z)}{f_q(z)} + \frac{1}{z}\frac{f_q'(z)}{f_q(z)} - \frac{1}{z^2}(q^2 z^2 + m^2 + d^2/4)\right] = 0 \ ,
\end{equation}
which can be solved by choosing $f_q(z) = K_\nu(|q|z)$ or $I_\nu(|q|z)$ where
\begin{equation}
	\nu = \sqrt{\left(\frac{d}{2}\right)^2+m^2} \ .
\end{equation}
To find $c(q)$ we need to substitute the ansatz into the inhomogeneous boundary condition. Choosing the solution which is regular in the interior we find that the boundary condition is satisfied if $c(q) \propto |q|^{d/2}/q$ \cite{Fujita:2011fp}. Setting $d = 4$ and $m^2 = 0$, for example, we obtain
\begin{equation}
	\phi(y,z) \propto \frac{  \left(2 y^3+3 y z^2\right)}{\left(y^2+z^2\right)^{3/2}} = 2 - \frac{3z^4}{4y^4} + \mathcal{O}(z^6/y^6) \ .
\end{equation}
The vacuum expectation value can then be read off from coefficient of the normalizable term
\begin{equation}
	\langle \mathcal{O}_4(y) \rangle \propto \frac{1}{y^4},
\end{equation}
which is of the form required by the unbroken subgroup of the conformal group, namely,
\begin{equation}
	\langle \mathcal{O}_\Delta(y) \rangle = \frac{c}{y^\Delta}.
\end{equation}

\subsection{Two-point function}
Unlike the holographic interface CFT \cite{Aharony:2003qf,DeWolfe:2001pq}, the holographic BCFT two-point function is not of the form $1/x^{2\Delta}$ even when $a=0$ \cite{Alishahiha:2011rg}. If we insert a boundary at $z=z_\ast$ then we need to impose the Neumann boundary condition at $y=z\sinh\rho_\ast$. For convenience we will choose $\rho_\ast = 0$ so that the boundary condition is simply
\begin{equation}
	\left.\partial_y \phi\right|_{y=0} =0 \ .
\end{equation}
Substituting the ansatz $\phi = z^{d/2}f(z)h(y) e^{- i \vec{\omega} \cdot \vec{x}}$ we find that the boundary condition  fixes $h(y) = e^{-iqy} + e^{iqy}$ and the general solution is thus of the form
\begin{equation}
	\phi(\vec{x},y,z) = z^{d/2}\int d^{d-1}\vec{\omega} \, dq \, (e^{-iq y} + e^{iq y})e^{-i\vec{\omega}\cdot\vec{x}}k^\nu K_\nu(k z)\phi_{(0)}(\vec{\omega},q) \ .
\end{equation}
Since $(e^{-iqy}+e^{iqy})K_\nu(kz)$ is an even function of $q$, the integral over $q$ projects out the even part of $\phi_{(0)}(\vec{\omega},q)$. Hence the only constraint on $\phi_{(0)}(\vec{\omega},q)$ is that it is itself an even function of $q$,
\begin{equation}
	\phi_{(0)}(\vec{\omega},q) = \phi_{(0)}(\vec{\omega},-q) \ .
\end{equation}
Fourier transforming, we have
\begin{align}
	\phi_{(0)}(\vec{x},y) 
		& = \int d^{d-1}\vec{\omega}\, dq \, e^{-i(qy + \vec{\omega}\cdot\vec{x} )}\phi_{(0)}(\vec{\omega},q) \ , \\
		& = \int d^{d-1}\vec{\omega}\, dq \cos(qy) e^{-i\vec{\omega}\cdot\vec{x}}\phi_{(0)}(\vec{\omega},q) \ , \\
		& = \frac{1}{2}\int d^{d-1}\vec{\omega}\, dq (e^{iqy}+e^{-iqy}) e^{-i\vec{\omega}\cdot\vec{x}}\phi_{(0)}(\vec{\omega},q) \ .
\end{align}
Inverting the Fourier transform we then find
\begin{equation}
	\phi_{(0)}(\vec{\omega},q) = \frac{1}{2(2\pi)^d}\int d^{d-1}\vec{x} \, dy (e^{iqy}+e^{-iqy})e^{i\vec{\omega}\cdot\vec{x}}\phi_{(0)}(\vec{x},y) \ ,
\end{equation}
which is automatically invariant under $q\to-q$ for any $\phi_{(0)}(\vec{x},y)$. Substituting back we obtain
\begin{align}
	\phi(\vec{x},y,z) 
		& = \frac{1}{2}\int d^{d-1}\vec{x}'dy' \bigl[K_\Delta(\vec{x},y;\vec{x}',y',z)+K_\Delta(\vec{x},-y;\vec{x}',y',z) + \nn\\
		& \quad +K_\Delta(\vec{x},y;\vec{x}',-y',z)+K_\Delta(\vec{x},-y;\vec{x}',-y',z)\bigr]\phi_{(0)}(\vec{x}',y') \ , \\
		& = \int d^{d-1}\vec{x}'dy' \bigl[K_\Delta(\vec{x},y;\vec{x}',y',z)+K_\Delta(\vec{x},-y;\vec{x}',y',z) \bigr]\phi_{(0)}(\vec{x}',y') \ , 
\end{align}
where
\begin{align}
	K_\Delta(\vec{x},y;\vec{x}',y',z)
		& = z^{d/2} \int \frac{d^{d-1}\vec{\omega} \,  dq}{(2\pi)^d} e^{iq(y-y')+i\vec{\omega}(\vec{x}-\vec{x}')} k^\nu K_\nu(k z) 
\end{align}
is the standard bulk-to-boundary propagator for an operator of dimension $\Delta = d/2+\nu$.
The two-point function is
\begin{equation}
	\langle \mathcal{O}(X_1)\mathcal{O}(X_2) \rangle = \frac{1}{|X_1-X_2|^{2\Delta}} + \frac{1}{|X_1-X_2^\ast|^{2\Delta}} \ ,
\end{equation}
where $X \equiv (\vec{x},y)$ and $X^\ast \equiv (\vec{x},-y)$. Setting $\vec{x}_2=\vec{0}$ without loss of generality we obtain
\begin{align}
	\langle \mathcal{O}(\vec{x},y_1)\mathcal{O}(\vec{0},y_2) \rangle
		& = \frac{1}{(\vec{x}^2 + (y_1-y_2)^2)^\Delta}+\frac{1}{(\vec{x}^2 + (y_1+y_2)^2)^\Delta} \ ,\\
		& = \frac{1}{(4y_1y_2)^\Delta}\left[\frac{1}{\xi^\Delta} + \frac{1}{(\xi+1)^\Delta}\right] \ , \label{e:BCFT}
\end{align}
which is of the correct form \cite{Cardy:1984bb,McAvity:1995zd} dictated by conformal invariance,
\begin{equation}
	\langle \mathcal{O}_1(\vec{x},y_1)\mathcal{O}_2(\vec{0},y_2) \rangle = \frac{F(\xi)}{y_1^{\Delta_1}y_2^{\Delta_2}} \ , \quad \quad \xi = \frac{\vec{x}^2 + (y_1-y_2)^2}{4y_1y_2} \ .
\end{equation}
The function $F(\xi)$, which is not fixed by conformal invariance alone, is determined by the AdS/CFT calculation.
If we take $y_2 \to 0$ then
\begin{equation}
	\langle \mathcal{O}(\vec{x},y)\mathcal{O}(\vec{0},0) \rangle = \frac{2}{(\vec{x}^2 + y^2)^\Delta} \ ,
\end{equation}
which is of the form fixed by $\mathrm{O}(1,4)$ invariance
\begin{equation}
	\langle \mathcal{O}_1(\vec{x},y_1)\mathcal{O}_2(\vec{0},0) \rangle \propto \frac{1}{y_1^{\Delta_1-\Delta_2}(\vec{x}^2 + y_1^2)^{\Delta_2}} \ .
\end{equation}
Repeating the calculation for the Dirichlet boundary condition we obtain
\begin{align}
	\langle \mathcal{O}(X_1)\mathcal{O}(X_2) \rangle
		& = \frac{1}{|X_1-X_2|^{2\Delta}} - \frac{1}{|X_1-X_2^\ast|^{2\Delta}} \  \\
		& = \frac{1}{(4y_1y_2)^\Delta}\left[\frac{1}{\xi^\Delta} - \frac{1}{(\xi+1)^\Delta}\right] \ . \label{e:Dirichlet}
\end{align}
Note that, unlike in the case of Neumann boundary conditions, this two-point function vanishes in the limit $y_2 \to 0$.
\subsection{Two-point function in AdS slicing}
If we allow $\rho_\ast \neq 0$ then we encounter a difficulty because the boundary condition now mixes $y$ derivatives with $z$ derivatives on $Q$
\begin{equation}
	\partial_y\phi|_{y = z\tan\theta} = \cot\theta \partial_z\phi|_{y = z\tan\theta} \ .
\end{equation}
It is thus more natural to work in the slicing of AdS$_{d+1}$ by AdS$_{d}$, where the boundary condition is replaced by $\partial_\rho\phi|_{\rho_\ast} = 0$. The metric in these coordinates is given by
\begin{equation}
	ds^2 = d\rho^2 + \cosh^2\rho \, ds^2_{\mathbb{H}_d}.
\end{equation}  
We will mainly focus on the example of a marginal operator in three dimensions, but similar results hold for any $d$ and $\Delta$. As shown in appendix \ref{app:eigen}, the bulk-to-boundary propagator (assuming $d$ odd) is given in these coordinates by
\begin{equation}
	K(\rho,x;x') =
		\frac{2i}{V_{\mathbb{S}^d}(d-1)} \int_0^\infty dp f_p(\rho) p \, C^{(d-1)/2}_{-(d-1)/2 +ip}(\cosh \ell) \ ,
\end{equation}
where $f_p(\rho)$ is some linear combination of
\begin{equation}
	f(\rho)
		= (\sech\rho)^{d/2}\left\{ P_{-1/2 + ip}^{\nu}(\tanh\rho) , \quad Q_{-1/2 + ip}^{\nu}(\tanh\rho)\right\} \ , \label{conicalfunctions}
\end{equation}
to be fixed by boundary conditions.
Given a marginal operator we should, according to the AdS/CFT correspondence, consider a $m^2 = 0$ scalar field in AdS$_4$.  
We obtain (assuming $d=3$) the following asymptotics for the conical functions in \eqref{conicalfunctions} (the asymptotic boundary is at $\rho = - \infty$) ,
\begin{align}
(\sech\rho)^{d/2} P_{-1/2 + ip}^{\nu}(\tanh\rho)
	& = \sqrt{\frac{2}{\pi }} \cosh(p \pi )-2 \left(1+p^2\right) \sqrt{\frac{2}{\pi }} \cosh(p \pi ) e^{2\rho} \nonumber \\
	& \ \ \ \ \ \ \ \ \ \ \ +\frac{8}{3} p \left(1+p^2\right) \sqrt{\frac{2}{\pi }} \sinh(p \pi ) e^{3\rho} + \mathcal{O}(e^{4\rho}) \ ,\\
(\sech\rho)^{d/2}Q_{-1/2 + ip}^{\nu}(\tanh\rho)
	& = -i \sqrt{\frac{\pi }{2}} \sinh(p \pi )+i \left(1+p^2\right) \sqrt{2 \pi } \sinh(p \pi) e^{2\rho} \nonumber \\
	& \ \ \ \ \ \ \ \ \ \ \   -\frac{4}{3} i p \left(1+p^2\right) \sqrt{2 \pi } \cosh(p \pi) e^{3\rho} + \mathcal{O}(e^{4\rho}) \ .
\end{align}
Regularity in the interior ($\rho_\ast = \infty$) demands that we drop the Legendre-$P$ function and thus, according to the AdS/CFT dictionary, the bare two-point function is given by
\begin{equation}
	\langle\mathcal{O}_3(n)\mathcal{O}_3(n') \rangle_{\mathbb{H}_3} \propto \int_0^\infty dp \, p^2(1+p^2)\coth(\pi p) U_{-1+ip}(n\cdot n') \ .
\end{equation}
The right-hand side is clearly divergent, as is to be expected since we are dealing with the bare, rather than the renormalized correlator. We can gain considerable insight about this infinite expression with the help of the integral representation of the generalized Gegenbauer function \cite{durand}
\begin{align}
	C_{-(d-1)/2+i p}^{(d-1)/2}(z)
		& =  i(-1)^{(d-1)/2+1} 2^{-(d-1)/2}\frac{\sinh (\pi p)}{\pi}\int_{-\infty}^{\infty} d\beta (\cosh\beta + z)^{-(d-1)/2} e^{- i p \beta} \ .
\end{align}
Rotating the contour to the imaginary axis by defining $\sigma = i \beta$, and using the Mellin transformation, we obtain the following generating function
\begin{equation}
	p(\sigma) = (\cos \sigma + z)^{-(d-1)/2} = i (-2)^{(d-1)/2}\int_{0}^{\infty} dp \cosh(\sigma p) \, \frac{C_{-(d-1)/2+i p}^{(d-1)/2}(z)}{\sinh(\pi p)} \ .
\end{equation}
It is now possible to express the bare correlator as a linear combination of derivatives of $p(\sigma)$, thus extracting the finite part of the bare correlator
\begin{align}
	\langle\mathcal{O}_3(n)\mathcal{O}_3(n') \rangle_{\mathbb{H}_3}
		& \propto \left.\left(\frac{d^2p}{d\sigma^2}+ \frac{d^4p}{d\sigma^4}\right)\right|_{\sigma=\pi} = \frac{6}{(n\cdot n'-1)^3} \ .
\end{align}
We thus see that the finite piece agrees with the expectations from conformal invariance.

For the BCFT we obtain a linear combination of $Q$ and $P$ Legendre functions determined by the boundary condition $f'_p(\rho_\ast) = 0$. In particular,
\begin{equation}
	f_p(\rho) = (\sech\rho)^{d/2} \left[ Q_{-1/2+ip}^{\nu}(\tanh\rho) + b_p P_{-1/2+ip}^{\nu}(\tanh\rho) \right] \ ,
\end{equation}
where
\begin{equation}
	b_p 
		= -\frac{(1+2ip-2\nu)Q_{1/2+ip}^\nu(\tanh\rho_\ast) + (d-1-2ip)Q_{-1/2+ip}^\nu(\tanh\rho_\ast)\tanh\rho_\ast}{(1+2ip-2\nu)P_{1/2+ip}^\nu(\tanh\rho_\ast) + (d-1-2ip)P_{-1/2+ip}^\nu(\tanh\rho_\ast)\tanh\rho_\ast} \ .
\end{equation}
Notice that for $\rho_\ast \to 0$ we obtain
\begin{equation}
	b_p = -\frac{Q_{1/2+ip}^\nu(0)}{P_{1/2+ip}^\nu(0)} \ ,
\end{equation}
while for $\rho_\ast \to \infty$ we have $b_p \to 0$ and we recover the formula for a pure CFT. 

Choosing $d =3$, $m^2 = 0$, $\rho_\ast = 0$ and using our Gaussian test function $f(z) = e^{-z}$, we obtain
\begin{equation}
	\frac{i}{2\pi^2} \int_{0}^\infty dp \, p \, U_{-1 +ip}(1) g(p) f(p)^\ast \simeq -1.19857 \ , \quad \quad g(p) = \frac{\pi \coth(\pi p) + 2 i b_p}{2i b_p \coth(\pi p) + \pi} \ .
\end{equation}
In order to evaluate the RHS of the distribution formula, we need to conformally map the BCFT two-point function \eqref{e:BCFT} to the hyperboloid. This can be achieved by identifying $y > 0$ with the Poincar\'{e} radial coordinate of $\mathbb{H}_d$. It then follows that
\begin{equation}
	\cosh\ell - 1 = \frac{(z-z')^2 + (\vec{x}-\vec{x}')^2}{2zz'} = 2\xi \ ,
\end{equation}
and thus
\begin{align}
	\langle \mathcal{O}(n)\mathcal{O}(n') \rangle_{\mathbb{H}_d}
		& = \frac{1}{2^\Delta} \left[\frac{1}{(\cosh\ell - 1)^\Delta} + \frac{1}{(\cosh\ell + 1)^\Delta}\right] \ .
\end{align}
Subtracting divergences and smearing with the test function over the hyperboloid we obtain
{\small\begin{align}
	\int_{\mathbb{H}_3} dn \langle\mathcal{O}_3(n)\mathcal{O}_3(n') \rangle_{\mathbb{H}_3}^{\rm ren} f(n \cdot n')
		& = \lim_{\eta\to 0}\left[4\pi\int_{1+\eta}^\infty dz(z^2-1)^{1/2}e^{-z}\frac{1}{2^3}\left[\frac{1}{(z-1)^3}+\frac{1}{(z+1)^3}\right]-\frac{\pi\sqrt{2}}{3\eta^{3/2}}e^{-1} + \frac{3\pi}{2\sqrt{2\eta}}e^{-1}\right] \notag \\
		& \simeq 1.19857 \ .
\end{align}}
In the case of a Dirichlet boundary condition at $\rho_\ast = 0$, the two-point function \eqref{e:Dirichlet} expressed in terms of the geodesic distance is
\begin{align}
	\langle \mathcal{O}(n)\mathcal{O}(n') \rangle_{\mathbb{H}_d}
		& = \frac{1}{2^\Delta} \left[\frac{1}{(\cosh\ell - 1)^\Delta} - \frac{1}{(\cosh\ell + 1)^\Delta}\right] \ ,
\end{align}
and
\begin{equation}
	b_p = -\frac{Q_{-1/2+ip}^\nu(0)}{P_{-1/2+ip}^\nu(0)} \ .
\end{equation}
In this case
{\small\begin{align}
	\int_{\mathbb{H}_3} dn \langle\mathcal{O}_3(n)\mathcal{O}_3(n') \rangle_{\mathbb{H}_3}^{\rm ren} f(n \cdot n')
		& = \lim_{\eta\to 0}\left[4\pi\int_{1+\eta}^\infty dz(z^2-1)^{1/2}e^{-z}\frac{1}{2^3}\left[\frac{1}{(z-1)^3}-\frac{1}{(z+1)^3}\right]-\frac{\pi\sqrt{2}}{3\eta^{3/2}}e^{-1} + \frac{3\pi}{2\sqrt{2\eta}}e^{-1}\right] \notag \\
		& \simeq 1.1333 \  , \\
	\frac{i}{2\pi^2} \int_{0}^\infty dp \, p \, U_{-1 +ip}(1) g(p) f(p)^\ast 
		& \simeq -1.1333 \ .
\end{align}}

\section{Holographic pseudo-conformal CFT}\label{sec:pseudo}
The pseudo-conformal CFT can be regarded as a CFT with a spacelike boundary at future infinity. We begin with the AdS$_{d+1}$ metric \eqref{adsmetricp} and perform the coordinate transformation $z = (-\eta) \csch \rho$ and $t = \eta \coth\rho$. Then
\begin{align}
	ds^2
		& = d\rho^2 + \sinh^2\rho\frac{-d\eta^2 +d\vec{x}^2}{\eta^2} \  \\
		& = d\rho^2 + \sinh^2\rho \, ds^2_{dS_d} \ ,
\end{align}
where $\eta \in (-\infty,0)$ and $\rho \in (0,\infty)$. Unlike the AdS$_d$ slicing, this coordinate system only covers a subregion of the AdS$_{d+1}$ Poincar\'{e} patch. The subregion already has a boundary given by the light-cone at $\rho = 0$. Rather than choosing $Q$ to be this null boundary, however,we will instead fix $Q$ at some $\rho_\ast > 0$.

\begin{figure}[h]
\centering
\includegraphics[width=90mm]{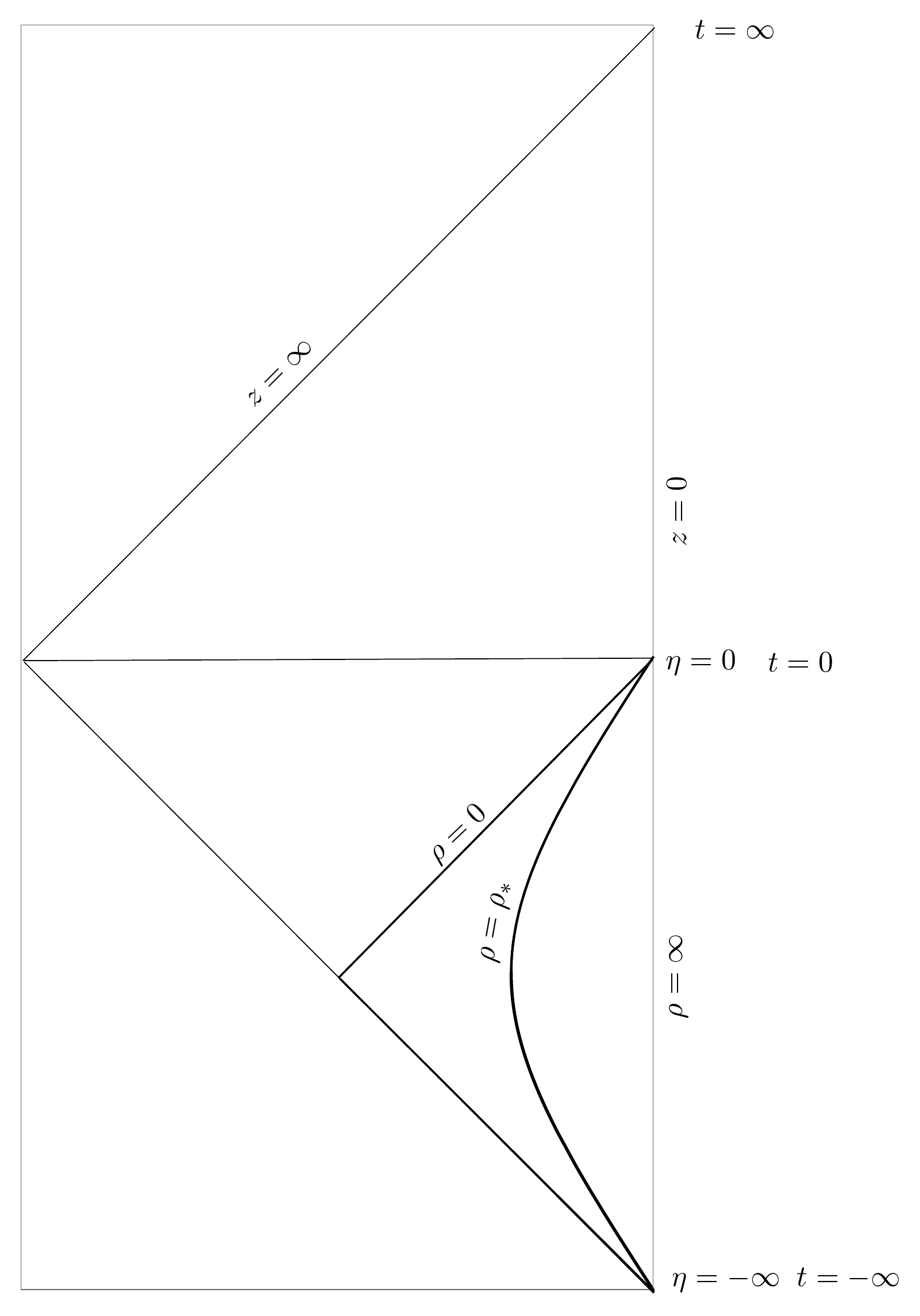}
\caption{Setup for the pseudo-conformal BCFT, showing the Poincar\'{e} patch covered by $z,t$, the region covered by the dS slice coordinates $\rho,\eta$, and the surface $Q$ at $\rho=\rho_\ast$ intersecting the $\rho=\infty$ boundary at $t=0$.}
\label{fig2}
\end{figure}

We expect that the resulting VEV will be of the form $1/(-t)^\Delta$ with a $\rho_\ast$-dependent coefficient which vanishes as $\rho_\ast \to 0$. A general $\rho = \rho_\ast$ surface is given in Poincar\'{e} coordinates by a worldline
\begin{equation}
	(-t) = z \cosh\rho_\ast \ ,
\end{equation}
which intersects the boundary on the spacelike surface $t = 0$, as in Figure \ref{fig2}. Let us define $\cosh\rho_\ast = \coth\phi$ so that the surface $Q$ is defined by $t = -z\coth\phi$. Consider the Lorentz boost
\begin{equation}
	\begin{pmatrix}\tilde{t} \\ \tilde{z}\end{pmatrix} = \begin{pmatrix} \cosh\phi & \sinh\phi \\ \sinh\phi & \cosh\phi \end{pmatrix}\begin{pmatrix}t \\ z\end{pmatrix} \ .
\end{equation}
Now the surface is defined by $\tilde{z} = 0$, and the metric in these coordinates is
\begin{equation}
	ds^2 = \frac{d\tilde{z}^2 - d\tilde{t}^2 + d\vec{x}^2}{z^2} \ .
\end{equation}
The boundary condition on $Q$ is now
\begin{align}
	\left.\partial_{\tilde{z}}\phi\right|_{t=-z\coth\phi} + \frac{a}{z}
		& = 0 \ , \\
	\left.(\sinh\phi \partial_t + \cosh\phi \partial_z )\phi\right|_{t=-z\coth\phi}  + \frac{a}{z}
		& = 0 \ .
\end{align}

\subsection{One-point function}
For $0 < a <\infty$ we choose the ansatz to be
\begin{equation}
	\phi(t,z) = z^{d/2}\int_{-\infty}^{\infty} dq \, f_q(z) c(q) e^{i q t} \ .
\end{equation}
Substituting into the scalar equation we obtain
\begin{equation}
	z^{d/2+2}\int_{-\infty}^{\infty} dq\, f_q(z)c(q)e^{iqt}\left[\frac{f_q''(z)}{f_q(z)} + \frac{1}{z}\frac{f_q'(z)}{f_q(z)} - \frac{1}{z^2}(-q^2 z^2 + m^2 + d^2/4)\right] = 0 \ .
\end{equation}
Demanding that the terms in square brackets vanish we find that $f_q(z)$ should be a linear combination of Bessel functions, which have the following asymptotic behavior 
\begin{align}
	\mathbb{Y}_\nu(|q|z)
		& = z^{ \nu } \left(-\frac{2^{-\nu } |q|^{\nu } \cos(\pi  \nu) \Gamma(-\nu )}{\pi }+\mathcal{O}(z^2)\right)+z^{-\nu}\left(-\frac{2^{\nu } |q|^{-\nu } \Gamma(\nu)}{\pi }+\mathcal{O}(z^2)\right) \ , \\
	J_\nu(|q|z)
		& = z^{\nu } \left(\frac{2^{-\nu } |q|^{\nu }}{\Gamma(1+\nu)}+\mathcal{O}(z^2)\right) \ .
\end{align}
These asymptotic expansions suggest that in order to interpret the scalar field configuration as a spontaneously generated VEV, we should choose $f_q(z) = J_\nu(|q|z)$. Substituting this into the boundary condition we then obtain
\begin{align}
	\int_{-\infty}^{\infty}dq \, \left[iq \sinh\phi + \frac{1}{z}\left(\frac{d}{2} -1 - q\frac{c'(q)}{c(q)} + iq z\coth\phi  \right)\cosh\phi\right]J_\nu(|q|z)c(q)e^{-iq z\coth\phi} &	= -\frac{a}{z^{d/2+1}} \ ,
\end{align}
where we have replaced $z\partial_z$ by $q\partial q$ and integrated by parts. Notice that for large arguments the Bessel function is oscillating rather than decaying exponentially
\begin{equation}
	J_\nu(z) \sim \sqrt{\frac{2}{\pi z}}\cos \left(z - \nu\pi/2 - \pi/4\right) \ .
\end{equation}
The function $c(q)$ cannot be a power law $|q|^{\alpha}$ because this would imply
\begin{align}
	\frac{1}{z^2}\int_{-\infty}^{\infty}d\tilde{q} \, \left[i\tilde{q} \sinh\phi + \left(\frac{d}{2} -1 - \alpha + i\tilde{q}\coth\phi  \right)\cosh\phi\right]J_\nu(\tilde{q})c(\tilde{q}/z)e^{-i\tilde{q}\coth\phi} &	= -\frac{a}{z^{d/2+1}} \ ,
\end{align}
and then $\alpha = d/2-1$, which would cause the LHS to diverge. On the other hand, if we choose $c(q)$ to be a regulated delta function
\begin{equation}
	c(q) = \frac{1}{2\alpha}e^{-\alpha |q|} \ ,
\end{equation}
then assuming $d=4$ and $m^2=0$ we obtain
\begin{align}
	\phi(y,z)
		& = z^2\int_{-\infty}^\infty dq \, J_\nu(|q|z)\frac{1}{2\alpha}e^{-\alpha |q|}e^{i q t} \ , \\
		& = \left(-\frac{3}{4 t^4}+\mathcal{O}(\alpha^2)\right) z^4+\left(-\frac{5}{4 t^6}+\mathcal{O}(\alpha^2)\right) z^6+\mathcal{O}(z)^8 \ .
\end{align}
The correctly normalized scalar is
\begin{align}
	\phi(y,z)
		& = \frac{a \left(2-2 \sqrt{1-x^2}+x^2 \left(-3+2 \sqrt{1-x^2}\right)\right) \csch^4\phi \sech(2 \phi )}{3 \left(1-x^2\right)^{3/2}} \ ,
\end{align}
where $x = z/(-t)$. By direct substitution it can be shown that this solves both the equation of motion and the boundary condition. For the same reason as in the timelike BCFT case, the dual operator acquires a VEV
\begin{equation}
	\langle \mathcal{O}_4(t)\rangle \propto \frac{1}{t^4}.
\end{equation}
\subsection{Two-point function}
In the Euclidean signature the space dS$_d$ continues to $\mathbb{S}^d$ and the  wave equation in AdS$_{d+1}$ sliced by $\mathbb{S}^d$ is solved by
\begin{align}
	\phi(\rho,\hat{n})
		& = \sum_{l,\mathbf{m}} f_l(\rho) \mathbb{Y}_{l\mathbf{m}}(\hat{n}) \phi_{(0)l\mathbf{m}} \  \\
		& = \frac{1}{V_{\mathbb{S}^d}(d-1)}\int d\Omega_{d}'\sum_{l}(2l+d-1) f_l(\rho) C_l^{(d-1)/2}(\hat{n}\cdot\hat{n}') \phi_{(0)}(\hat{n}') \ ,
\end{align}
where 
\begin{align}
	f_l(\rho)
		& = (\sinh\rho)^{(1-d)/2} \left\{P_{-1/2+\nu}^{(1-d)/2 - l}(\cosh\rho), \quad Q_{-1/2+\nu}^{(1-d)/2 - l}(\cosh\rho)\right\} \ .
\end{align}
The asymptotics of the ring functions can be found from the relations (see sec.~3.13 of \cite{bateman})
\begin{align}
	P^\mu_{-1/2+\nu}(\cosh\rho) 
		& = \frac{2^{2\mu}}{\Gamma(1-\mu)}(1-e^{-2\rho})^{-\mu}e^{-(\nu + 1/2)\rho} F(1/2-\mu,1/2+\nu-\mu;1-2\mu;1-e^{-2\rho}) \ , \\
	Q^\mu_{-1/2+\nu}(\cosh\rho) 
		& = \frac{\pi^{1/2}e^{i\mu\pi}\Gamma(1/2+\nu+\mu)}{\Gamma(1+\nu)}(1-e^{-2\rho})^{\mu}e^{-(\nu + 1/2)\rho} F(1/2+\mu,1/2+\nu+\mu;1+\nu;e^{-2\rho})  \ .
\end{align}
Setting $d=3$ and $m^2 = 0$, 
we obtain
\begin{align}
	(\sinh\rho)^{(1-d)/2} P_{-1/2+\nu}^{(1-d)/2 - l}(\cosh\rho)
		& = \frac{1}{\Gamma(3+l)}-\frac{2 (1+l)^2e^{-2\rho}}{\Gamma(3+l)}+\frac{8 e^{-3\rho}}{3 \Gamma(l)} + \mathcal{O}(e^{-4\rho}) \ , \\
	(\sinh\rho)^{(1-d)/2} Q_{-1/2+\nu}^{(1-d)/2 - l}(\cosh\rho)
		& = -\frac{8}{3} (-1)^l \Gamma(1-l)e^{-3\rho} + \mathcal{O}(e^{-5\rho}) \ .
\end{align}
For a massless scalar in AdS$_{d+1}$ (assuming $d$ odd) there are similar expressions in which $3$ is replaced by $d$. It follows that the holographic two-point function on $\mathbb{S}^d$ is given by
\begin{equation}
	\langle \mathcal{O}(\vec{n})\mathcal{O}(\vec{n}')\rangle \propto \sum_{l=0}^\infty (2l+d-1)\frac{\Gamma(d+l)}{\Gamma(l)}C_l^{(d-1)/2}(\vec{n}'\cdot\vec{n}') \ .
\end{equation}
Comparing with \eqref{e:gl} we obtain
\begin{equation}
	\langle \mathcal{O}(\vec{n})\mathcal{O}(\vec{n}') \propto \frac{1}{(1-\cos\Theta)^{d}} \ ,
\end{equation}
which is the correct result for a marginal operator in $d$ dimensions.

The infinite sums defining our holographic two-point functions do not converge. They can be regulated, however, by using the following generating function for Gegenbauer polynomials
\begin{equation}
	p(t) = \frac{1}{(1-2tx + t^2)^{\nu}} = \sum_{l=0}^\infty C_l^\nu(x) t^l \ .
\end{equation}
Let us check this explicitly for $d=3$,
\begin{align}
	\langle \mathcal{O}(\vec{n})\mathcal{O}(\vec{n}')\rangle 
		& \propto \sum_{l=1}^\infty l(l+1)^2 (l+2)U_l(\cos\Theta) \ , \\
		 & = p''''(1) + 10 p'''(1)+24p''(1) + 12p'(1) \ . \\
		& = \frac{3}{(1-x)^3} \ .
\end{align}
In the general situation with a boundary in the bulk, we require a linear combination
\begin{equation}
	f_l(\rho) = (\sech\rho)^{1-d/2} \left[P_{-1/2+\nu}^{(1-d)/2-l}(\cosh\rho) +  b_l Q_{-1/2+\nu}^{(1-d)/2-l}(\cosh\rho) \right] \ ,
\end{equation}
where the coefficient $b_l$ is determined by imposing either a Neumann ($f_l'(\rho_\ast) = 0$) or Dirichlet ($f_l(\rho_\ast) = 0$) condition at $\rho = \rho_\ast$. As we move $\rho_\ast \to \infty$ the boundary disappears, $b_l \to 0$ and we recover the pure CFT.

\section{Discussion}
We have developed the formalism that relates the spectral decomposition of correlation functions to the renormalized correlation functions in position space. 
In highly symmetric situations the spacetime representation of the two-point function can be deduced from spacetime symmetries. This is true of both the pure CFT on flat and curved backgrounds as well as CFTs with spacelike or timelike boundaries. We have checked in all these cases that both representations are related by integral transforms. 

In other situations such as the pseudo-conformal CFT where the exact spacetime form of two-point function is not known, our formalism allows it to be computed implicitly from a knowledge of the spectral representation obtained via holography. It would be interesting to try to compare the results of that calculation with the low-momentum expansion of the two-point functions computed from effective field theory considerations in \cite{Hinterbichler:2011qk}.

So far our calculations have been restricted to Euclidean signature, although it would be interesting to extend them to the Lorentzian signature, in which the foliation of AdS$_{d+1}$ by $\mathbb{S}^d$ becomes a foliation by $d$-dimensional de Sitter slices.

\vspace{.5cm}
\noindent
{\bf Acknowledgments}:  The authors would like to thank Steven Gubser, Paul McFadden and Massimo Porrati for helpful discussions and correspondence.
Research at Perimeter Institute is supported by the Government of Canada through Industry Canada and by the Province of Ontario through the Ministry of Economic Development and Innovation. This work was made possible in part through the support of a grant from the John Templeton Foundation.  The opinions expressed in this publication are those of the authors and do not necessarily reflect the views of the John Templeton Foundation (K.H.).  The work of M.T. was supported in part by the US Department of Energy, and J.S. and M.T. are supported in part by NASA ATP grant NNX11AI95G.

\appendix

\section{Green's functions and propagators}

Here we collect some results about scalar propagators and Green's functions in various maximally symmetric spaces.

\subsection{Bulk-to-boundary propagator in anti-de Sitter slicing}\label{app:eigen}
The wave equation in the $\mathbb{H}_d$ slicing of $\mathbb{H}_{d+1}$ is
\begin{equation}
	\partial_\rho^2 \phi + d \tanh\rho \, \partial_\rho \phi + \sech^2\rho \, \nabla_{\mathbb{H}_d}^2 \phi = m^2\phi \ .
\end{equation}
Separating variables by writing $\phi = f(\rho)g(x)$ we obtain
\begin{equation}
	\cosh^2\rho\left[\frac{f''(\rho)}{f(\rho)} +  d \tanh\rho\frac{f'(\rho)}{f(\rho)} - m^2 \right] + \frac{\nabla_{\mathbb{H}_d}^2 g(x)}{g(x)}  = 0 \ ,
\end{equation}
and thus we need to solve the following eigenvalue problem
\begin{align}
	+l(l+d-1)
		& =\frac{\nabla_{\mathbb{H}_d}^2 g(x)}{g(x)} \ ,\\
	0
		& =\frac{f''(\rho)}{f(\rho)} +  d \tanh\rho\frac{f'(\rho)}{f(\rho)} - m^2 + \lambda\sech^2\rho \ .
\end{align}
Consider the flat slicing of $\mathbb{H}_d$,
\begin{equation}
	ds_{\mathbb{H}_d}^2 = \frac{d\eta^2 + d\vec{x}^2}{\eta^2} \ .
\end{equation}
There are two branches of solutions depending on whether $\lambda \equiv l(l+d-1)$ is above or below the Breitenlohner-Freedman bound \cite{Breitenlohner:1982jf,Breitenlohner:1982bm} for AdS$_d$
\begin{equation}
	\lambda_{\rm BF} = -\left(\frac{d-1}{2}\right)^2 \ .
\end{equation}
We will focus on the range $\lambda < \lambda_{\rm BF}$ since this is required to obtain a complete set. Defining the complex angular momentum
\begin{equation}
	l = - \frac{d-1}{2} + i p \ , \quad p > 0 \ ,
\end{equation}
we have
\begin{equation}
	\lambda = \lambda_{\rm BF} - p^2 \ ,
\end{equation}
and the general solutions are given by the linear combinations
\begin{align}
	g(\vec{x},\eta)
		& = \eta^{(d-1)/2}  \left\{ K_{ip} (k\eta), \quad I_{ip}(k\eta) \right\}e^{-i\vec{k}\cdot\vec{x}} \ , \\
	f(\rho)
		& = (\sech\rho)^{d/2}\left\{ P_{-1/2 + ip}^{\nu}(\tanh\rho) , \quad Q_{-1/2 + ip}^{\nu}(\tanh\rho)\right\} \ ,
\end{align}
where $\nu = \sqrt{d^2/4 + m^2}$. The modified Bessel functions of imaginary order (MacDonald functions) satisfy the Sturm-Liouville differential identity
\begin{equation}
\frac{d}{dx}\left[x\frac{d}{dx}K_{ip}(x)\right] - xK_{ip}(x) + \frac{p^2}{x}K_{ip}(x) = 0 \ ,
\end{equation}
where the weight function $w(x) = 1/x$ is positive for $x > 0$. This ensures that they obey the orthogonality relations \cite{Pass09}
\begin{align}
	\frac{2}{\pi^2} p \sinh(\pi p) \int_0^\infty dx \frac{K_{ip}(x)K_{ip'}(x)}{x} \ ,
		& = \delta(p-p') \\
	\frac{2}{\pi^2 x}\int_0^\infty dp \, p \sinh(\pi p) K_{ip}(x) K_{ip}(y) 
		& = \delta(x-y) \ .
\end{align}
The normalized wavefunctions on $\mathbb{H}_d$ satisfy
\begin{equation}
	\int d^d x\sqrt{g_{\mathbb{H}_d}} \psi_{k,p}(x)\psi_{k',p'}(x)^\ast = \delta^{d-1}(\vec{k}-\vec{k}')\delta(p-p') \ ,
\end{equation}
and are given by
\begin{equation}
	\psi_{k,p}(\vec{x},\eta) = \eta^{(d-1)/2} \sqrt{\frac{p \sinh(\pi p)}{2^{d-2}\pi^{d+1}}}  K_{ip}(k \eta)e^{- i \vec{k}\cdot\vec{x}} \ ,
\end{equation}
which agrees with \cite{Grosche:1987dq} for $d=2$. Notice that these wavefunctions vanish at $\eta = \infty$ because $K_{i\rho}(k\eta)\sim \sqrt{\frac{\pi}{2k\eta}}e^{-k\eta}$, while at small $\eta$ they behave as $\psi_{k,p} \sim \eta^{(d-1)/2-ip}$, which vanishes as $\eta \to 0$. The $I_{i\rho}(k\eta)$ are not permissible wavefunctions because while they vanish at $\eta = 0$ they exponentially diverge as $\eta \to \infty$.

Now consider the spherical slicing of $\mathbb{H}_d$. The equation defining $g$ is now
\begin{equation}
	\partial_r^2 g + (d-1) \coth r \, \partial_r g(r) +\csch^2 r \nabla_{\mathbb{S}^{d-1}}^2 g = \lambda g \ .
\end{equation}
Separating variables as $g = Y(\hat{n})R(r)$ we then need to solve the following eigenvalue problem
\begin{align}
	\frac{\nabla_{\mathbb{S}^{d-1}}^2Y}{Y}
		& = - l(l+d-2) \ , \\
	0
		& =R''(r) + (d-1)\coth(r) R'(r) - \left[l(l+d-2)\csch^2 r + \lambda\right]R(r) \ .
\end{align}
We find that $Y$ is a hyper-spherical harmonic on $\mathbb{S}^{d-1}$ and the general solution for $R(r)$ (assuming $\lambda < \lambda_{\rm BF}$) is
\begin{equation}
	R(r) = (\sinh r)^{(2-d)/2}\left\{P_{-1/2+ip}^{(2-d)/2-l}(\cosh r),\quad Q_{-1/2+ip}^{(2-d)/2-l}(\cosh r)\right\} \ .
\end{equation}
The associated Legendre functions $P_{-1/2+i\rho}^\nu(x)$, with complex degree $-1/2+ip$, are called conical functions and satisfy the completeness relations 
\begin{align}
	\int_0^\infty dp \left|\frac{\Gamma(ip-\nu+1/2)}{\Gamma(ip)} \right|^2 P^\nu_{-1/2+ip}(x)P^\nu_{-1/2+ip}(y)
	 & = \delta(x-y) \ , \\
	\left|\frac{\Gamma(ip-\nu+1/2)}{\Gamma(ip)} \right|^2\int_1^\infty dx \,  P^\nu_{-1/2+ip}(x)P^\nu_{-1/2+ip'}(x)
	 & = \delta(p-p') \ .
\end{align}
We therefore find that the normalized wavefunctions in the $\mathbb{S}^{d-1}$ slicing of $\mathbb{H}_d$ are \cite{Grosche:1987de}
\begin{equation}
	\psi_{p,l\mathbf{m}}(r,\Omega) = \frac{\Gamma((d-1)/2 + ip + l)}{\Gamma(ip)} (\sinh r)^{(2-d)/2}P_{-1/2+ip}^{(2-d)/2-l}(\cosh r)\mathbb{Y}_{l\mathbf{m}}(\Omega_{d-1}) \ .
\end{equation}
We conclude that the general solution of the massive scalar wave equation in AdS$_{d+1}$ is
\begin{align}
	\phi(\rho,x) 
		& = \sum_{l,\mathbf{m}}\int dp \, f_p(\rho) \psi_{p,l\mathbf{m}}(x) \phi_{(0)p,l\mathbf{m}} \ , \\
		& = \sum_{l,\mathbf{m}}\int dp \, f_p(\rho) \psi_{p,l\mathbf{m}}(r,\Omega) \left[\int dV_{\mathbb{H}_d}' \psi_{p,l\mathbf{m}}^\ast(x')\phi_{(0)}(x')\right] \ , \\
		& =  \int dV_{\mathbb{H}_d}'\int dp \, f_p(\rho) \left[\sum_{l,\mathbf{m}}\psi_{p,l\mathbf{m}}(x) \psi_{p,l\mathbf{m}}^\ast(x')\right]\phi_{(0)}(x') \ , \\
		& = \int dV_{\mathbb{H}_d}'K(\rho,x;x')\phi_{(0)}(x')\ ,
\end{align}
where we have used \eqref{e:addition} and defined the bulk-to-boundary propagator
\begin{align}
	K(\rho,x;x')
		& = \frac{1}{2\pi}(2\pi\sinh \ell)^{(2-d)/2} \int dp \, f_p(\rho) \left|\frac{\Gamma((d-1)/2+ip )}{\Gamma(ip)} \right|^2 P_{-1/2+ip}^{(2-d)/2}(\cosh \ell) \ .
\end{align}

\subsection{Scalar Green's functions}

Here we review the scalar Green's functions on maximally symmetric spaces.

\subsubsection{Sphere $\mathbb{S}^{d+1}$}
We consider the scalar field action
\begin{equation}
	S = \frac{1}{2}\int_{\mathbb{S}^{d+1}} d^{d+1}x \sqrt{g} \left[(\nabla\phi)^2 + m^2 \phi^2\right] \ ,\label{scalarsphereaction}
\end{equation}
The standard round metric on the sphere is
\begin{equation}
	ds^2 = d\theta^2 + \sin^2\theta d\Omega_{d}^2 \ ,
\end{equation}
where $\theta\in (0,\pi)$ and the wave equation for a scalar of mass $m$ is 
\begin{equation}
	\left[\partial_\theta^2 + d\cot\theta \partial_\theta + \csc^2\theta \nabla_{\mathbb{S}^{d}}^2 - m^2\right]\phi = 0 \ .
\end{equation}
The Green's function when acted upon by the above differential operator gives a unit normalized delta function source. We can use rotational invariance to move the delta function source to $\theta = 0$ so that the Green's function only depends on the $\theta$ coordinate, and thus the $\nabla_{\mathbb{S}^{d}}^2$ term can be set to zero. Defining $z = \frac{1}{2}(1+\cos\theta)$ we have $dz/d\theta = - [z(1-z)]^{-1/2}$ and thus the equation of motion away from coincident points ($z\neq 1$) becomes
\begin{equation}
	z(1-z)G''(z) + (d+1)(1/2-z)G'(z) - m^2G(z) = 0 \ .
\end{equation}
Comparing with the hypergeometric equation
\begin{equation}
	z(1-z)F''(z) + [c-(a+b+1)z]F'(z) - abF(z) = 0 
\end{equation}
we obtain $c = (d+1)/2$, $ab = m^2$ and $a+b=d$. The last two relations give $a(d-a)=m^2$. The hypergeometric function equation has three singular points at $z=0,1,\infty$. The linearly independent solutions around each of these points are
\begin{align}
	&z=0: && F(a,b;c;z)\ ,  && z^{1-c}F(1+a-c,1+b-c;2-c;z) \ , \notag\\
	&z=1: && F(a,b;1+a+b-c;1-z) \ ,  && (1-z)^{c-a-b}F(c-a,c-b;1+c-a-b;1-z) \ , \notag \\
	&z=\infty: && z^{-a}F(a,1+a-c;1+a-b;z^{-1})\ , && z^{-b}F(b,1+b-c;1+b-a;z^{-1}) \ .
\end{align}
For the sphere $z\in [0,1]$, and we expect a singularity at $\theta = 0$ ($z=1$) but want to avoid a singularity at $\theta = \pi$ ($z = 0$). Smoothness at $\theta=\pi$ implies that we discard the second solution around $z=0$ and the first solution around $z=1$. Moreover, we can discard the second solution around $z=1$ because it is singular at $\theta=\pi$.  The solution is thus the original hypergeometric function,
\begin{equation}
	\bar{G}(\theta) =  F\left(\delta, d-\delta;\frac{d+1}{2};\frac{1}{2}(1+\cos\theta)\right) \ ,
\end{equation}
where bar indicates that we have dropped an overall normalization factor. The parameter $\delta$ (not to be confused with $\Delta$) is chosen to be the larger root of the quadratic equation $\delta(d-\delta)=m^2$; namely,
\begin{equation}
	\delta = \frac{d}{2}+ \sqrt{\left(\frac{d}{2}\right)^2 - m^2} \ .
\end{equation}
This choice is without loss of generality because of the hypergeometric identity
\begin{equation}
	F(a,b;c;z)=F(b,a;c;z) \ .
\end{equation}
This Green functions behaves in the expected way for a conformally coupled scalar on $\mathbb{S}^{d+1}$ (which has\footnote{Recall that a conformally coupled scalar has $m^2=\frac{d-1}{4d} R$ and $R = (d+1)d$ is the scalar curvature for the unit $d+1$-sphere.} $m^2 = (d+1)(d-1)/4$ and $\delta = (d+1)/2$),
\begin{align}
	\bar{G}(\theta) = F\left(\frac{d+1}{2}, \frac{d-1}{2};\frac{d+1}{2};\frac{1}{2}(1+\cos\theta)\right) = \left(\frac{2}{1-\cos\theta}\right)^{(d-1)/2} \ .
\end{align}
For a massless scalar $m^2=0$, there is subtlety due to the shift symmetry of the action and the resultant divergence over the zero mode causes the propagator to be divergent in the massless limit.  If we interpret the shift symmetry as a gauge symmetry, the two-point function turns out to be the coefficient of $m^2$ in the Taylor expansion of the normalized Green function $G(\theta)$ (see \cite{Folacci:1992xc} for details). In the case of a massless scalar on $\mathbb{S}^4$ we obtain 
\begin{equation}
	\bar{G}(\theta) = 1 + \frac{1}{3}\left[\frac{1}{2}\frac{z}{1-z}-\log(1-z)\right]m^2 + \mathcal{O}(m^4) \ ,
\end{equation}
and the divergent normalization factor ($\sim 1/m^2$) selects the second term.
 
It is also interesting to express the Green's function in terms of the heat Kernel on $\mathbb{S}^{d+1}$,
\begin{align}
	K(\hat{n},\hat{n}';t)
		& = \frac{1}{V_{\mathbb{S}^{d+1}}d}\sum_{l=0}^\infty (2l + d)C_\ell^{d/2}(\vec{n}\cdot\vec{n}') e^{- t\ell(\ell + d)} \ .
\end{align}
The Green's function is given by the Laplace transform of the heat kernel which provides a spectral decomposition analogous to \eqref{e:spheredecomp},
\begin{align}
	G(\hat{n},\hat{n}';m^2)
		& = \int_{0}^\infty dt \, e^{-m^2 t}K(\vec{n},\vec{n}';t) \ \\
		& = \frac{1}{V_{\mathbb{S}^{d+1}}d}\sum_{l=0}^\infty  \frac{2l+d}{l(l+d)+m^2}C_\ell^{d/2}(\vec{n}\cdot\vec{n}') \  \\
		& = \frac{1}{V_{\mathbb{S}^{d+1}}d} \frac{\pi}{\sin(\pi \nu)}C_\nu^{d/2}(-\vec{n}\cdot\vec{n}') \ ,
\end{align}
where $\nu$ satisfies $m^2=-\nu(\nu+d)$. We can see that this agrees with the previous calculation by making use of the representation of the Gegenbauer function in terms of a hypergeometric function 
\begin{equation}
	C_\nu^{d/z}(z) = C_\nu^{d/z}(1) F\left(-\nu,\nu+d;\frac{d+1}{2}; \frac{1-z}{2}\right) \ , \quad \quad C_\nu^{d/2}(1) = \frac{\Gamma(\nu+d)}{\Gamma(\nu+1)\Gamma(d)} \ .
\end{equation}

\subsubsection{$\mathbb{H}_{d+1}$}
The wave equation on $\mathbb{H}_{d+1}$ can be obtained from that on $\mathbb{S}^{d+1}$ by analytically continuing the polar coordinate $\theta = ir$ and simultaneously flipping the sign of the curvature, which flips $m^2 \to - m^2$, yielding
\begin{equation}
	\left[\partial_r^2  + d\coth r \, \partial_r +\csch^2 r \nabla_{\mathbb{S}^{d-1}}^2 - m^2 \right]\phi = 0 \ .
\end{equation}
Here $r\in (0,\infty)$.
Defining $z = \frac{1}{2}(1+\cosh r)$, $z\in (0,\infty)$, we have $dz/dr = [z(z-1)]^{-1/2}$ and the equation of motion away from coincident points becomes
\begin{equation}
	z(1-z)G''(z) + (d+1)(1/2-z)G'(z) + m^2G(z) = 0.
\end{equation}
The unnormalized Green's function obtained by analytical continuation from the sphere is given by
\begin{equation}
	\bar{G}_{\rm E}(\ell) = F\left(\Delta, d-\Delta;\frac{d+1}{2};\frac{1}{2}(1+\cosh\ell)\right) \ , \quad \quad \Delta = \frac{d}{2} + \sqrt{\left(\frac{d}{2}\right)^2 + m^2} \ ,
\end{equation}
where $\ell$ is the usual geodesic distance and $\Delta$ is the larger of the two roots $\Delta_\pm$ of the quadratic equation $\Delta(d-\Delta)=-m^2$. For large arguments the hypergeometric function has the following asymptotics (assuming $a-b$ is non-integer)
\begin{equation}
	F(a,b;c;z) = z^{-a}\left[\lambda_1 + \mathcal{O}(z^{-1})\right] + z^{-b}\left[\lambda_2 + \mathcal{O}(z^{-1})\right] \ ,
\end{equation}
and thus the above Green's function behaves asymptotically for large $\ell$ as
\begin{equation}
	G_{\rm E}(\ell) \sim Ae^{-\Delta_+ \ell} + Be^{-\Delta_-\ell} \ .
\end{equation}
The second term means that this Green's function is finite only if $m^2 \leq 0$.
If we consider the first solution of the hypergeometric equation around $z=\infty$ we find
\begin{equation}\label{e:adsgreen1}
	\bar{G}(\ell) = u^{\Delta}F\left(\Delta,\Delta+\frac{1-d}{2}; 2\Delta-d + 1; u\right) \ , \quad \quad u = \frac{2}{1+\cosh \ell} \ ,
\end{equation}
which behaves for large $\ell$ as $G(\ell) \sim e^{-\Delta \ell}$. The second solution around $z=\infty$ gives the same expression \eqref{e:adsgreen1} with $\Delta=\Delta_-$. We can use the hypergeometric identity (\cite{bateman} sec.~2.10, p.~109)
\begin{equation}
	F(a,b;c,u) = (1-u)^{-a}F\left(a,c-b;c;\frac{u}{u-1}\right) \ ,
\end{equation}
to express \eqref{e:adsgreen1} in the equivalent form
\begin{equation}\label{e:adsgreen5}
	\bar{G}(\ell) = (2v^{-1})^\Delta F\left(\Delta,\Delta+\frac{1-d}{2}; 2\Delta-d + 1; -2v^{-1}\right) \ , \quad \quad v=\frac{1}{2}(n_\mu-n'_\mu)(n_\nu-n'_\nu)\eta^{\mu\nu} = \cosh\ell-1 \ .
\end{equation}
We can also use the hypergeometric identity from sec.~2.1.5 of \cite{bateman} (p.~66)
\begin{equation}\label{e:identity}
	F(a,b;2b,u) = \left(1-\frac{u}{2}\right)^{-a}F\left(\frac{a}{2},\frac{a+1}{2};b+\frac{1}{2};\frac{u^2}{(2-u)^2}\right) \ ,
\end{equation}
which gives
\begin{equation}\label{e:adsgreen2}
	\bar{G}(\ell) = (2\xi)^{\Delta}F\left(\frac{\Delta}{2},\frac{\Delta+1}{2}; \Delta + 1-\frac{d}{2}; \xi^2\right) \ , \quad \quad \xi = \sech \ell \ .
\end{equation}
Remembering to remove the factor of $2^\Delta$, we fix notation be defining the `Feynman' Green's function to be
\begin{equation}
	\bar{G}_\Delta(\ell) =\xi^{\Delta}F\left(\frac{\Delta}{2},\frac{\Delta+1}{2}; \Delta + 1-\frac{d}{2}; \xi^2\right).
\end{equation}
If $m^2$ lies in the range
\begin{equation}
 -\left(\frac{d}{2}\right)^2 <	m^2 < -\left(\frac{d}{2}\right)^2 + 1 \ ,
\end{equation}
then the most general Green's function compatible with the AdS isometries is a linear combination
\begin{equation}
	\bar{G}(\ell) = \alpha \bar{G}_{\Delta_+}(\ell) + \beta \bar{G}_{\Delta_-}(\ell) \ .
\end{equation}
An important example is provided by a conformally coupled scalar on AdS$_{d+1}$, which has a mass given by
\begin{equation}
	m^2 = -\left(\frac{d}{2}\right)^2 + \frac{1}{4} \ ,
\end{equation}
so both $\Delta_\pm$ branches are allowed. In fact, conformal covariance of $\bar{G}(\ell)$ actually requires that both Green's functions appear in the linear combination $\alpha =(d-1)/2$, $\beta =1$. Let us see this explicitly for a conformally coupled scalar on AdS$_4$ which has $\Delta_+=2$ and $\Delta_-=1$,
\begin{equation}
	\bar{G}(\ell) = \alpha\frac{1}{\cosh^2\ell-1} + \beta \frac{\cosh \ell}{\cosh^2\ell-1} \ .
\end{equation} 
We notice that for $\alpha=\beta=1$, $\bar{G}(\ell)$ is proportional to the  Weyl transform of the CFT two-point function from flat space
\begin{equation}
	\frac{1}{\cosh^2\ell-1} + \frac{\cosh \ell}{\cosh^2\ell-1} = \frac{1}{\cosh \ell-1} \ .
\end{equation}
This boundary condition can be interpreted \cite{Porrati:2001db} as allowing the scalar energy to pass through the AdS$_4$ boundary into a second copy of AdS$_4$. Another interesting interpretation of this boundary condition is that it is precisely the one for which $\bar{G}(\ell)$ is proportional to the analytic continuation from the sphere $\bar{G}_{\rm E}(\ell)$. Using hypergeometric identifies it can be shown that \cite{Burgess:1984ti}
\begin{align}
	\bar{G}_\Delta(\ell) = A(d,\Delta) \left[\bar{G}_{\rm E}(\ell)+\tilde{G}_{\rm E}(\ell)\right]-B(d,\Delta) \left[\bar{G}_{\rm E}(\ell)-\tilde{G}_{\rm E}(\ell)\right] \ ,
\end{align}
where $\tilde{G}_{\rm E}(\ell)$ is related to $\bar{G}_{\rm E}(\ell)$ by taking $\cosh\ell \to - \cosh\ell$ and
\begin{align}
	A(d,\Delta)
		& = \frac{(-1)^{\Delta/2}\Gamma(\Delta-d/2+1)\Gamma\left(\frac{d+1-\Delta}{2}\right)}{2\Gamma\left(\Delta/2-d/2+1\right)\Gamma\left(\frac{d+1}{2}\right)} \ , \\
	B(d,\Delta)
		& = \frac{(-1)^{(\Delta+1)/2}\Gamma(\Delta-d/2+1)\Gamma\left(\frac{d-\Delta}{2}\right)}{2\Gamma\left(\Delta/2-d/2+1/2\right)\Gamma\left(\frac{d+1}{2}\right)} \ .
\end{align}
Demanding that the coefficient of $\tilde{G}_{\rm E}(\ell)$ vanishes leads to the boundary condition
\begin{equation}
	\alpha = -\frac{A(d,\Delta_-)+B(d,\Delta_-)}{A(d,\Delta_+)+B(d,\Delta_+)} \beta \ .
\end{equation}
Setting $\Delta_\pm = (d\pm 1)/2$ for a conformally coupled scalar we obtain $\alpha/\beta = (d-1)/2$.

Finally, let us note that there is a subtlety with using the $\Delta_-$ branch for a conformally coupled scalar in \eqref{e:adsgreen1} or \eqref{e:adsgreen5}. This is because $\Delta_- = (d-1)/2$ so the hypergeometric function becomes $F(\Delta,0,0,u)=1$. Instead one should first represent the hypergeometric function using \eqref{e:identity} and then take the limit $b \to 0$.

Let us derive these results from the sum over Brownian motions on the hyperboloid. The heat kernel on $\mathbb{H}_{d+1}$ is given by \cite{Grosche:1987de}
\begin{align}
	K(x,y;t)
		& = \frac{1}{2\pi}(2\pi\sinh \ell)^{(1-d)/2} \int_0^\infty dp\left|\frac{\Gamma(d/2+ip )}{\Gamma(ip)} \right|^2 P_{-1/2+ip}^{(1-d)/2}(\cosh \ell)e^{- t \left[p^2 + d^2/4\right]} \ .
\end{align}
For $d=1$ we have
\begin{equation}
	K(\ell;t) = \frac{1}{2\pi} \int_0^\infty dp \, p \tanh \pi p P_{-1/2+ip}(\cosh \ell)e^{- t \left[p^2 + 1/4\right]} \ ,
\end{equation}
and the associated Green's function is a ring function
\begin{align}
	G(\ell;m^2) 
		& = \int_0^\infty dt \, e^{-m^2 t}K(\ell;t) \  \\
		& = \frac{1}{2\pi} \int_0^\infty dp \, \frac{p \tanh \pi}{p^2 + \nu^2} P_{-1/2+ip}(\cosh \ell) \  \\
		& = \frac{1}{2\pi}Q_{-1/2+\nu}(\cosh \ell) \ ,
\end{align}
where now $\nu = \sqrt{1/4+m^2}$. The generalization to arbitrary $d$ (ignoring normalization) is given by
\begin{equation}\label{e:adsgreen3}
	\bar{G}(\ell;m^2) = (\sinh \ell)^{(1-d)/2}Q_{-1/2+\nu}^{(1-d)/2}(\cosh \ell) \ ,
\end{equation}
where $\nu = \sqrt{(d/2)^2+m^2}$. Let us now express this in terms of the hypergeometric function using sec.~3.2 of \cite{bateman} (p.~122), namely, 
{\small\begin{align}
	Q^\mu_\nu(z)
		& = e^{\mu i \pi} 2^{-\nu-1}\pi^{1/2}\frac{\Gamma(\nu+\mu+1)}{\Gamma(\nu+3/2)} z^{-\nu-\mu-1}(z^2 -1)^{\mu/2} F \left(\frac{\nu+\mu+2}{2},\frac{\nu+\mu+1}{2};\nu + \frac{3}{2};z^{-2}\right) \ , \\
		& = e^{\mu i \pi} 2^{-\nu-1}\pi^{1/2}\frac{\Gamma(\nu+\mu+1)}{\Gamma(\nu+3/2)} z^{-\nu-\mu-1}(z^2 -1)^{\mu/2}(1-z^{-2})^{-\mu} F \left(\frac{-\nu+\mu+1}{2},\frac{-\nu+\mu+2}{2};\nu + \frac{3}{2};z^{-2}\right) \ ,
\end{align}}
where in the second line we have used the Euler transformation
\begin{equation}
	F(a,b;c;z) = (1-z)^{c-a-b}F(c-a,c-b;c;z) \ , \quad \quad |z| < 1 \ ,
\end{equation}
which is applicable because $\sech l < 1$. Replacing $\mu \to (1-d)/2$ and $\nu \to -1/2+\nu$ we obtain \eqref{e:adsgreen2}.

\subsubsection{de Sitter}
The Bunch-Davies de Sitter two-point function can be obtained by analytic continuation from the sphere, $\theta = it + \pi/2$. Under this continuation the geodesic distance $\Theta$ defined by $\cos \Theta = \vec{n}\cdot\vec{n}' = \cos\theta_1\cos\theta_1 + \cos\alpha\sin\theta_1\sin\theta_2$ becomes
\begin{align}
	\cosh \ell = -\sinh t_1\sinh t_1 + \cos\alpha\cos t_1\cosh t_2 \ ,
\end{align}
where we recall that $\alpha$ is the angular separation in the sphere dS$_{d+1} \cong \mathbb{R}_t \times \mathbb{S}^d$.
The de Sitter Green's function can now be expressed in arbitrary coordinates by realizing that $\cosh \ell = \eta^{\mu\nu}n_\mu n_\nu'.$ where $\eta^{\mu\nu}$ is the (mostly plus) metric for $\mathbb{R}^{1,d+1}$ and $n$, $n'$ label two points on the single-sheeted hyperboloid defined by $n_\mu n^\mu = 1$.  The geodesic distance $l$  can be either real (for timelike separated points) or imaginary (for spacelike separations). In the flat slicing of de Sitter space we have
\begin{equation}
	\cosh \ell = \frac{\eta^2 + \eta'^2 - (\vec{x}-\vec{x}')^2}{2\eta \eta'} \ .
\end{equation}
The Bunch-Davies propagator has the following asymptotics for large $\ell$
\begin{equation}
	G_{\rm BD}(\ell) \sim Ae^{-\delta \ell} + Be^{-(d-\delta)\ell} \ .
\end{equation}
The observation that the Bunch-Davies propagator contains two asymptotic components has been used to argue that it cannot be defined as a sum over trajectories in de Sitter space. The alternative proposal is to take \cite{Polyakov:2007mm}
\begin{equation}
	G_{\rm dS}(n,n';m^2) = G_{\rm AdS}(n,n';-m^2) \sim Ae^{-\delta \ell} \ .
\end{equation}

\section{Differential regularization\label{diffregappend}}

We have seen in Section \eqref{flatspacesubsn} that the hard-cut off regulator introduces both power law and logarithmic divergences in the CFT two-point function on flat space. There should exist a regularization scheme in which only the logarithmic divergences appear, since the only unambiguous information present in the divergences is contained in the coefficients of these logarithms.  The differential regularization of \cite{Freedman:1991tk} is such a scheme.  Consider the two-point function of an operator of dimension $\Delta$. We begin by expressing $1/|x|^{2\Delta}$ in terms of an arbitrary number of Laplacians $\square^{k+1}$,
\begin{align}
	\frac{1}{|x|^{2\Delta}}
		& = \frac{1}{2^{2k+2}}\frac{\Gamma(\Delta - k-1)}{\Gamma(\Delta)}\frac{\Gamma(\Delta - k - d/2)}{\Gamma(\Delta-d/2+1)}\square^{k+1} \frac{1}{|x|^{2(\Delta - k-1)}}.
\end{align}
We notice that the coefficient diverges whenever $\Delta = d/2 + k$ where $k=0,1,2,\ldots$. Letting $\Delta = d/2+k+\epsilon$ we obtain
\begin{align}
	\frac{1}{|x|^{2\Delta}}
		& = \frac{1}{\epsilon(d+2\epsilon-2)}\frac{1}{2^{2k+1}}\frac{\Gamma(d/2+\epsilon)}{\Gamma(d/2+k+\epsilon)}\frac{\Gamma(1+\epsilon)}{\Gamma(k+1+\epsilon)}\square^k\left(\square\frac{1}{|x|^{d-2+2\epsilon}}\right) \ .
\end{align}
As $\epsilon \to 0$ two things happen; the object in parentheses approaches a delta function and the coefficient diverges.
Expanding in $\epsilon$ and keeping only the leading divergent term we find
\begin{equation}
	\frac{1}{|x|^{2\Delta}} \sim -V_{\mathbb{S}^{d-1}}\frac{1}{\epsilon}\frac{1}{2^{2k+1}}\frac{\Gamma(d/2)}{\Gamma(d/2+k)}\frac{1}{k!}\square^k\delta^{(d)}(x) \ .
\end{equation}
We therefore define the renormalized two-point function as
\begin{align}
	\langle \mathcal{O}_\Delta(x)\mathcal{O}_\Delta(0) \rangle
		& \equiv \lim_{\epsilon \to 0}\left[ \frac{1}{|x|^{2\Delta}} + V_{\mathbb{S}^{d-1}}\frac{\mu^{2\epsilon}}{\epsilon}\frac{1}{2^{2k+1}}\frac{\Gamma(d/2)}{\Gamma(d/2+k)}\frac{1}{k!}\square^k\delta^{(d)}(x)\right] \ ,
\end{align}
where we have introduced the mass scale $\mu$ to keep the equation dimensionally correct. As an example, consider $d=4$ and $\Delta = 2$ ($k =0$),
\begin{align}
	\langle \mathcal{O}_2(x)\mathcal{O}_2(0) \rangle 
		& = \lim_{\epsilon \to 0}\left[\frac{1}{x^4} + \frac{\pi^2 \mu^{2\epsilon}}{\epsilon}\delta^4(x)\right] \\
		& = \lim_{\epsilon \to 0}\left[\frac{1}{x^4} + \left(\frac{\pi^2}{\epsilon}+2\pi^2 \log \mu\right)\delta^4(x)\right]
\end{align}
so
\begin{equation}
	\mu \frac{\partial}{\partial \mu}\langle \mathcal{O}_2(x)\mathcal{O}_2(0) \rangle = 2\pi^2 \delta^4(x),
\end{equation}
 in agreement with \eqref{scalevariationf} obtained using the cutoff method.
 
We can re-express the delta function as a derivative to obtain an alternative expression for the two-point correlator
\begin{align}
	\langle \mathcal{O}_\Delta(x)\mathcal{O}_\Delta(0) \rangle
		& = -\frac{1}{d-2}\frac{1}{2^{2k+1}}\frac{\Gamma(d/2)}{\Gamma(d/2+k)}\frac{1}{k!}\square^{k+1}\frac{1}{|x|^{d-2}}\log (x^2\mu^2) \ .
\end{align}
To compute the Fourier transform we use\footnote{This formula requires differential regularization to make sense.} (see Eq.~(A.2) of \cite{Freedman:1991tk}),
\begin{equation}
	\int d^d x \, e^{i p x}\frac{1}{|x|^{d-2}}\log(\mu^2 x^2) = - \frac{4\pi^{d/2}}{\Gamma(d/2-1)} \frac{1}{p^2}\log(p^2/\bar{\mu}^2), \quad \quad \bar{\mu} \equiv 2\mu/\gamma.
\end{equation}
It follows that 
\begin{equation}
	\langle \mathcal{O}_\Delta(k)\mathcal{O}_\Delta(-k) \rangle \propto p^{2k}\log (p^2/\bar{\mu}^2).
\end{equation}

\end{document}